\documentclass[english]{emulateapj} \pdfoutput=1 \usepackage[]{fontenc}
\usepackage[latin1]{inputenc} \usepackage{amsmath} \usepackage{graphicx}
\usepackage{amssymb} \usepackage{color}
 \newcommand{\del}{\nabla}

\begin{document}

\title{Magnetically-levitating accretion disks around supermassive black holes}

\author{Evghenii Gaburov\altaffilmark{1,*},  Anders Johansen\altaffilmark{2}
and Yuri Levin\altaffilmark{3,4}}

\altaffiltext{1}{Centre for Interdisciplinary Exploration and Research in
Astrophysics, Northwestern University, 2131 Tech Drive, Evanston, IL 60208,
USA} \altaffiltext{2}{Lund Observatory, Department of Astronomy and Theoretical
Physics, Lund University, Box 43, 221 00 Lund, Sweden} \altaffiltext{3}{School
of Physics and Monash Center for Astrophysics, Monash University, Clayton, VIC
3800, Australia} \altaffiltext{4}{Leiden University, Leiden Observatory, Niels
Bohrweg 2, 2300 RA, Leiden, the Netherlands} \altaffiltext{*}{Hubble Fellow}

\begin{abstract} In this paper we report on the formation of
magnetically-levitating accretion disks around supermassive black holes. The
structure of these disks is calculated by numerically modelling tidal
disruption of magnetized interstellar gas clouds. We find that the resulting
disks are entirely supported by the pressure of the magnetic fields against the
component of gravitational force directed perpendicular to the disks. The
magnetic field shows ordered large-scale geometry that remains stable for the
duration of our numerical experiments extending over 10\% of the disk lifetime.
Strong magnetic pressure allows high accretion rate and inhibits disk
fragmentation.  This in combination with the repeated feeding of magnetized
molecular clouds to a supermassive black hole yields a possible solution to the
long-standing puzzle of black hole growth in the centres of galaxies.
\end{abstract}

% \maketitle

\section{Introduction}\label{sect:introduction}  

It is believed the growth of supermassive black holes (SMBHs) in centres of
galaxies is enabled by gas accretion from surrounding disks
\citep{1969Natur.223..690L} which have been observed with increasing precision
by modern telescopes \citep{1995Natur.373..127M,2004Natur.429...47J}. In the
early theoretical work \citep{1969Natur.223..690L,1973A&A....24..337S} it has
been suggested that that magnetic stresses play an important role in driving
the accretion by enabling the outward  angular-momentum transport through the
disk.  This suggestion has been put on a firm theoretical footing by
\cite{1991ApJ...376..214B} discovery of the importance of magnetorotational
instability (MRI) in astrophysical disks, and by the subsequent work, that
demonstrated the ability of MRI to build and maintain substantial magnetic
stresses inside the disk
\citep{1995ApJ...446..741B,1996ApJ...463..656S,2006ApJ...640..901H,2010ApJ...713...52D}.
%eg:add
 All of the numerical studies to date have demonstrated MRI-generated
magnetic stresses which are associated with the sub-thermal magnetic fields in
the disk mid-plane.  
%sub-dominant relative to the  gas pressure in the disk mid-plane.

One of the central unresolved issues of feeding SMBHs has been the tendency of
all modelled extended gaseous disks to clump due to their self-gravity
\citep{1980SvAL....6..357K,1987Natur.329..810S,1990Natur.345..679S,2003MNRAS.339..937G,
2009ApJ...704..281R}. Such choking of the accretion flow is a major obstacle in
SMBH growth.  It has been conjectured
\citep{1990ApJ...350..295S,2000ApJ...532L..67M,2003A&A...407..403P,2006PASJ...58..193M,
2007MNRAS.375.1070B, 2009ApJ...697...16O} that in some astrophysical disks
magnetic stresses may become dominant relative to the mid-plane gas pressure,
and that these disks may effectively resist fragmentation.  In this paper we
investigate the formation of accretion disks by performing numerical
simulations of collisions between magnetized gas clouds and a black hole. It
has been suggested that such collisions may be responsible for feeding the
supermassive black holes at the centers of galaxies
\citep{2007MNRAS.377L..25K,2008ApJ...683L..37W,2012ApJ...750L..38W} and that it
may have lead to the formation of the stellar disc in our own Galactic Center
\citep{,1998MNRAS.294...35S,2003ApJ...590L..33L,2006ApJ...643.1011P,
2008Sci...321.1060B, 2009MNRAS.394..191H}.  We find that the resulting disks
are completely dominated by the magnetic field pressure, and display high
accretion rates due to the Maxwell stress associated with the large-scale
magnetic field the structure of which remains stable over the duration of the
simulation. The Toomre-$Q$ factors of these naturally-formed
magnetically-levitating accretion disks (MLAD) indicate their stability to
gravitational fragmentation.  Therefore, MLADs represent a new class of
accretion-disk solutions which may play an important role in feeding the
supermassive black holes.

\section{Simulations setup}\label{sect:models}

We model a collision between a magnetized gas cloud and a SMBH using a new
moving-mesh ideal MHD scheme (see Appendix). We choose an equation of state
$P_{\rm gas}=c_{\rm s}^2 \rho$, where $c_{\rm s}=0.03\,v_{\rm K}$; here $v_{\rm
K}$ is Keplerian velocity around a SMBH. The temperature in this setup is
$T=1.63\times10^4\,{\rm K}\, (0.1\,{\rm pc}/R)$, which, in absence of magnetic
fields, would produce disks with $H_0/R=0.03$. In the presence of magnetic
fields the effective scale-height is modified by the magnetic pressure,
$H=H_0\,\sqrt{1+\beta^{-1}}$, where $\beta^{-1}=P_{\rm{m}}/P_{\rm{g}}$ is the
ratio of magnetic to gas pressures.  In order to isolate the effects of
magnetic fields on the disk formation process, we ignore effects of the gas
self-gravity in our calculations.

\begin{figure*}[] \begin{center}
\includegraphics[scale=0.27]{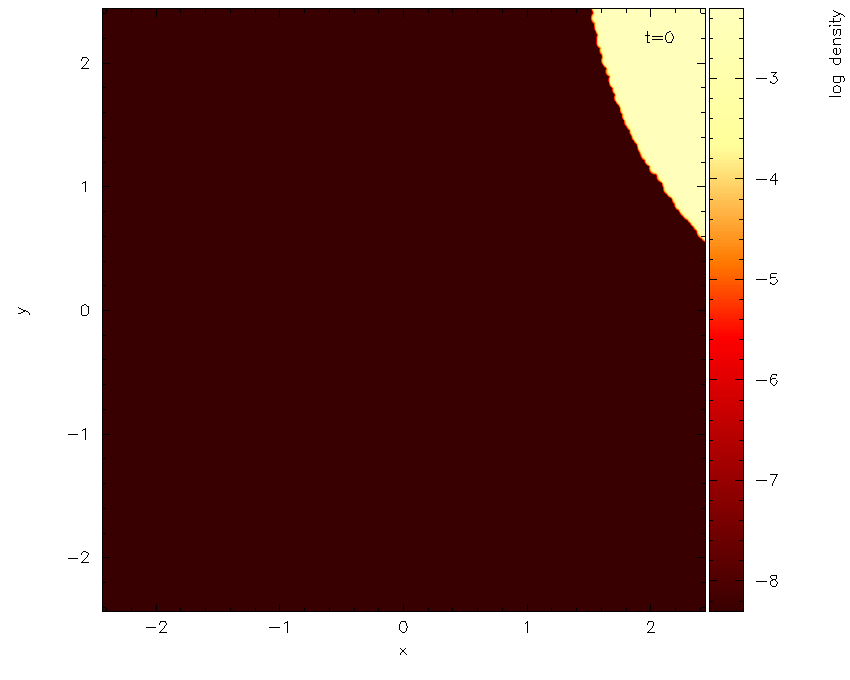}
\includegraphics[scale=0.27]{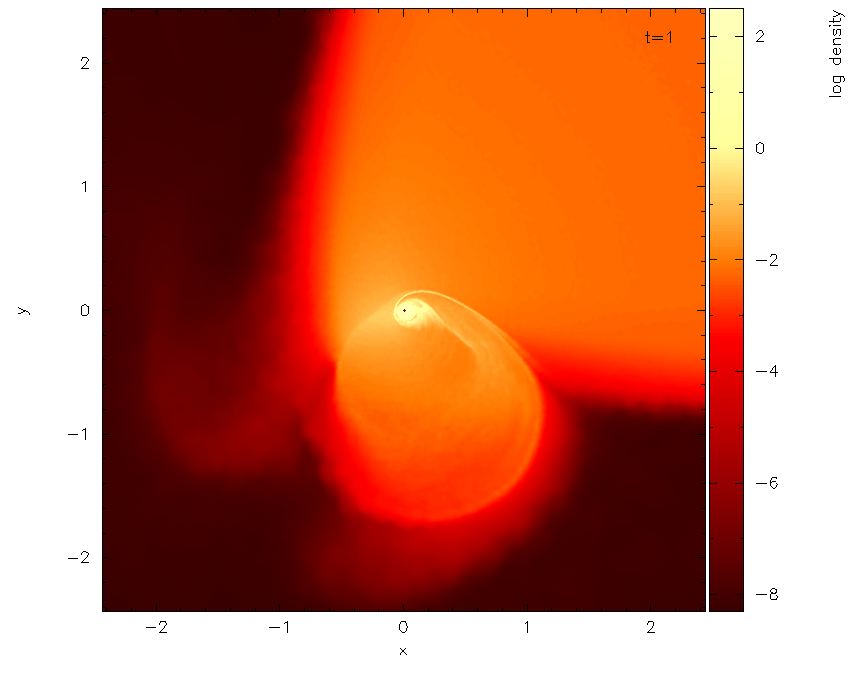}
\includegraphics[scale=0.27]{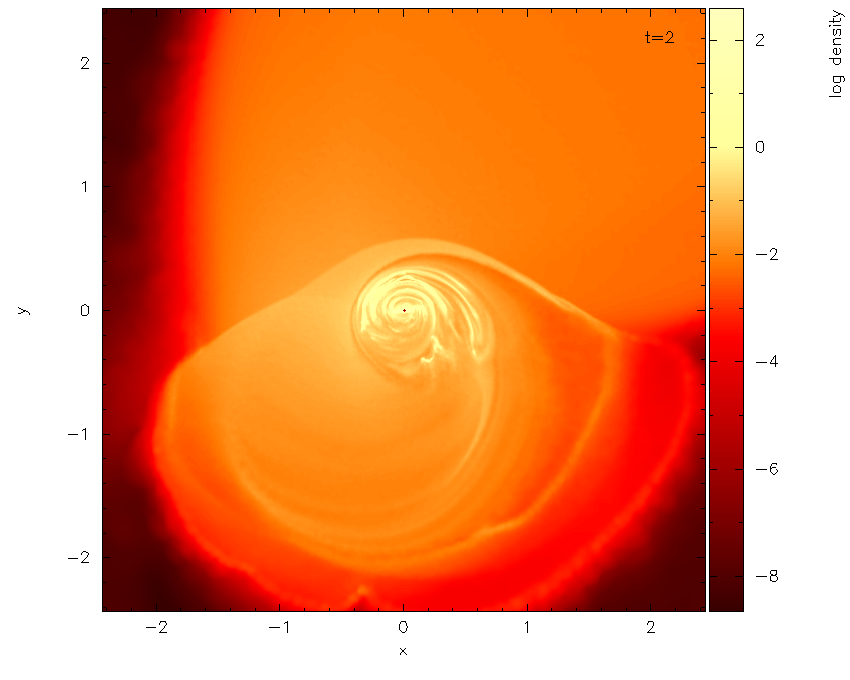}
\includegraphics[scale=0.27]{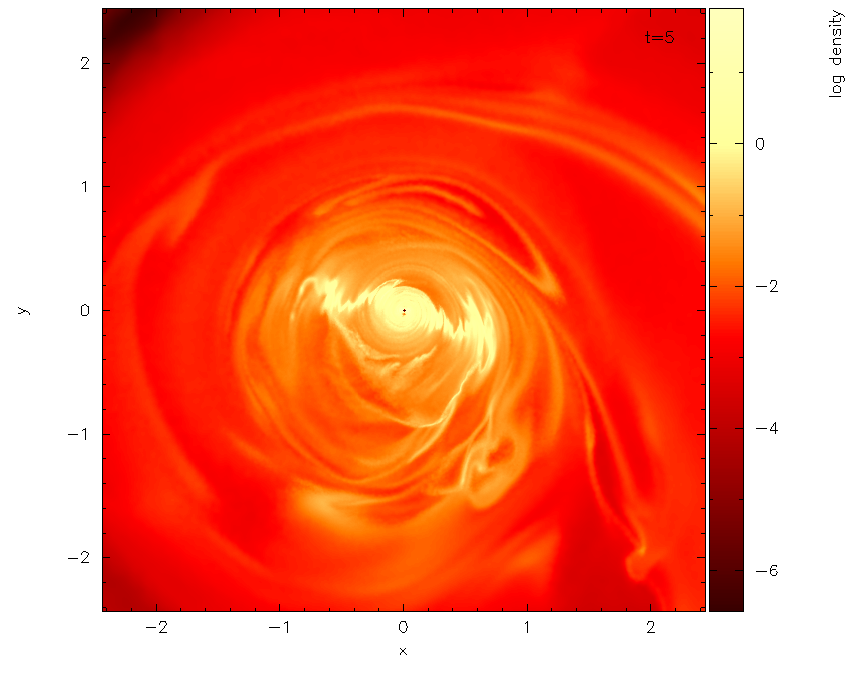} \end{center}
\caption{Snapshots of density structure in the $XY$ plane at different times
for V01 model. The times are shown in units of $0.047$ million years, which
corresponds to 0.0, 0.047, 0.096 and  0.240 million years for top-left,
top-right, bottom-left and bottom-right panels respectively. The unit of length
is a parsec, and unit of density is $4.1\cdot10^6\,$cm$^{-3}\,m_u$.}
\label{fig:V01dens} \end{figure*} \begin{figure*}[] \begin{center}
\includegraphics[scale=0.27]{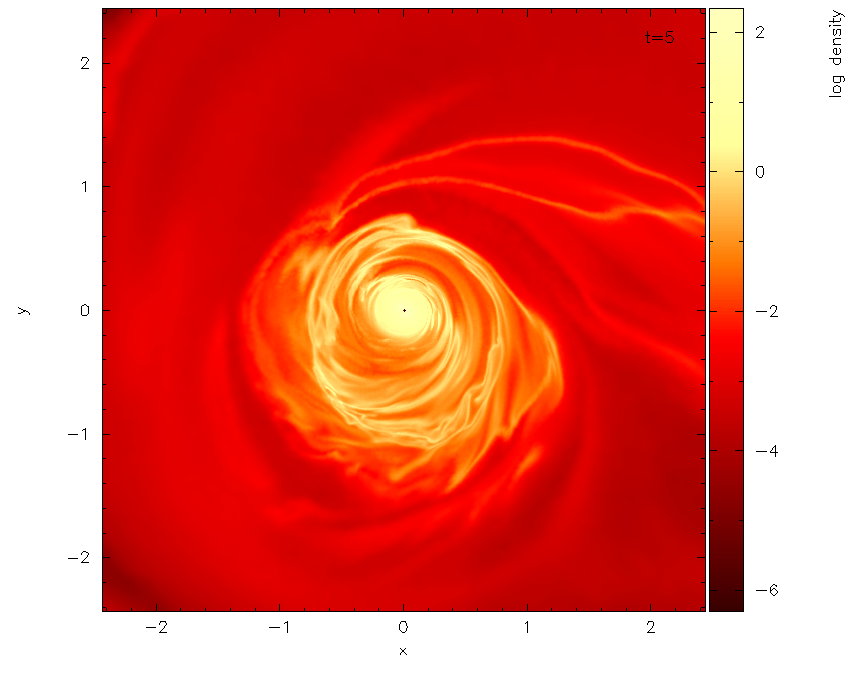}
\includegraphics[scale=0.27]{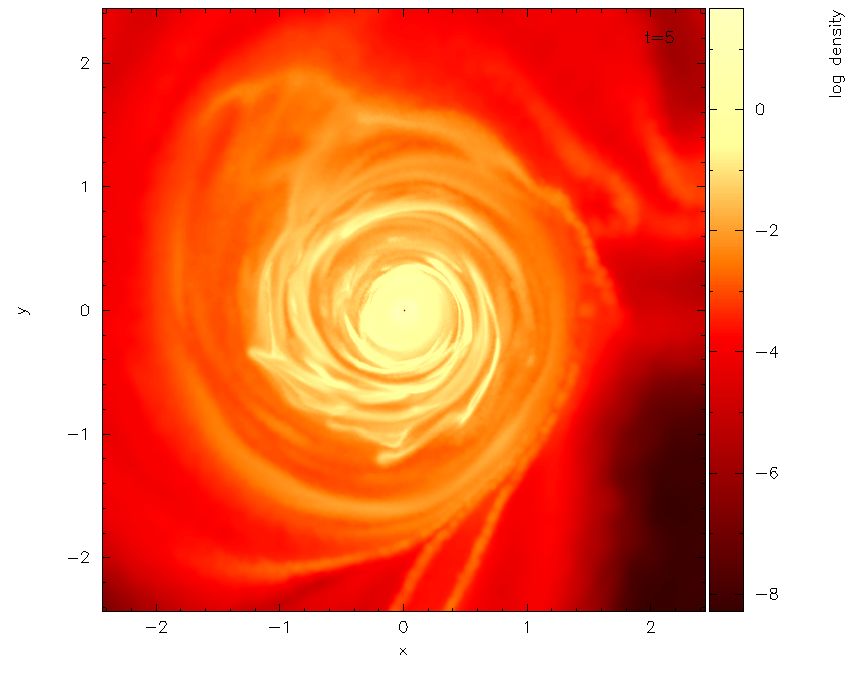}
\includegraphics[scale=0.27]{img/snap80}
\includegraphics[scale=0.27]{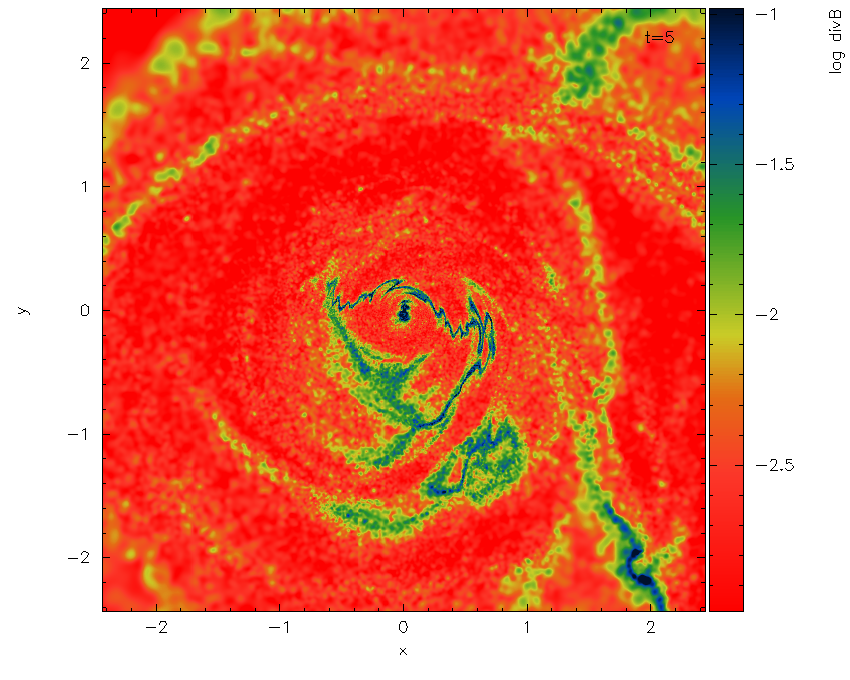} \end{center} \caption{Density
snapshot in the $XY$ plane at the of V02 (top left-panel) and C01 (top
right-panel) simulations. As in Fig.\,\ref{fig:V01dens}, the unit of length is
a parsec, and density is shown in units of $4.1\cdot10^6\,m_u/$cm$^3$.  The
bottom left and right panels show density profile in $XY$ plane at the end of
V01 simulation and the associated divergence error, divB $= |\del\cdot{\bf
B}|h/|{\bf B}|$ where $h$ is cell size, respectively.} \label{fig:V02C01dens}
\end{figure*}

\begin{figure*}[] \begin{center}
\includegraphics[scale=0.16]{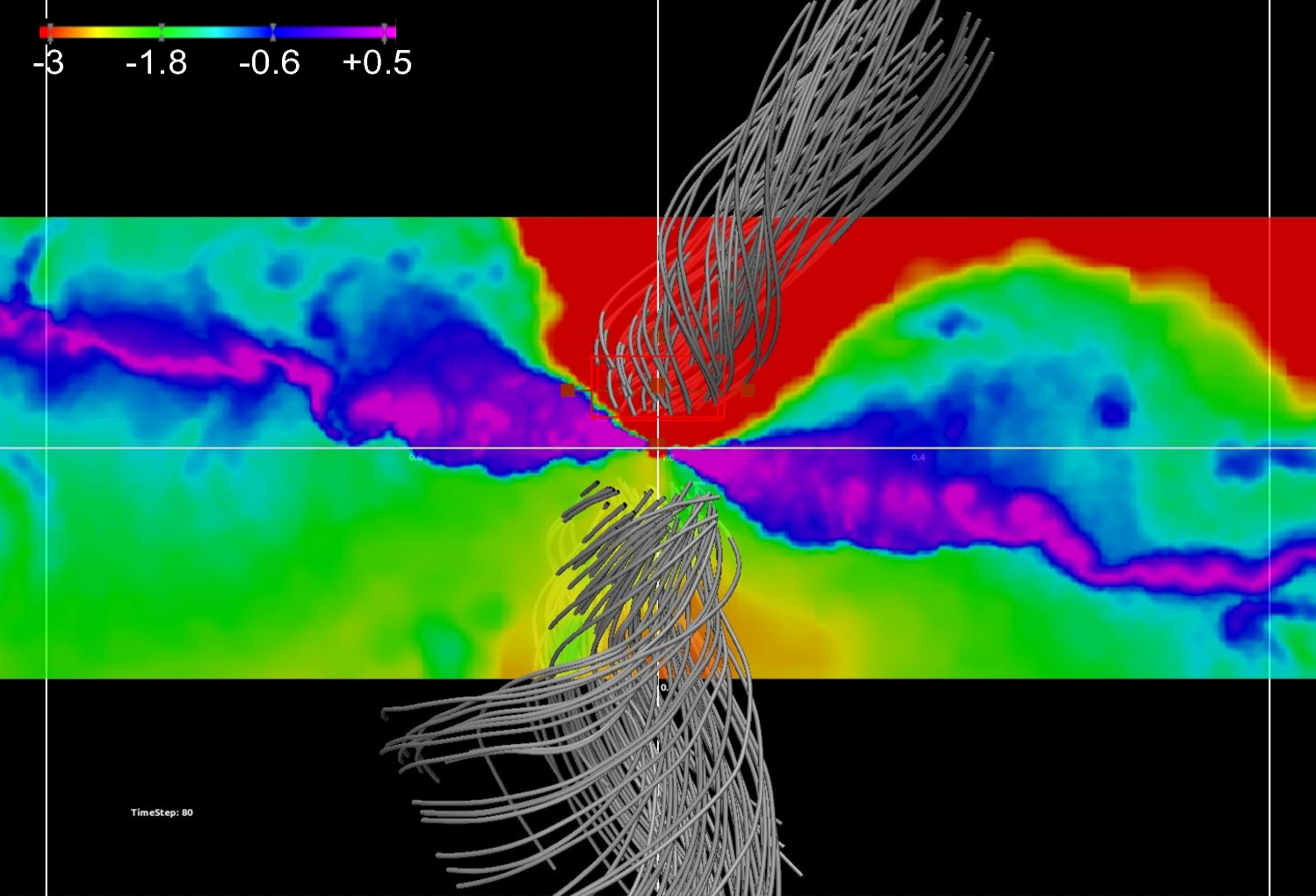}
\includegraphics[scale=0.16]{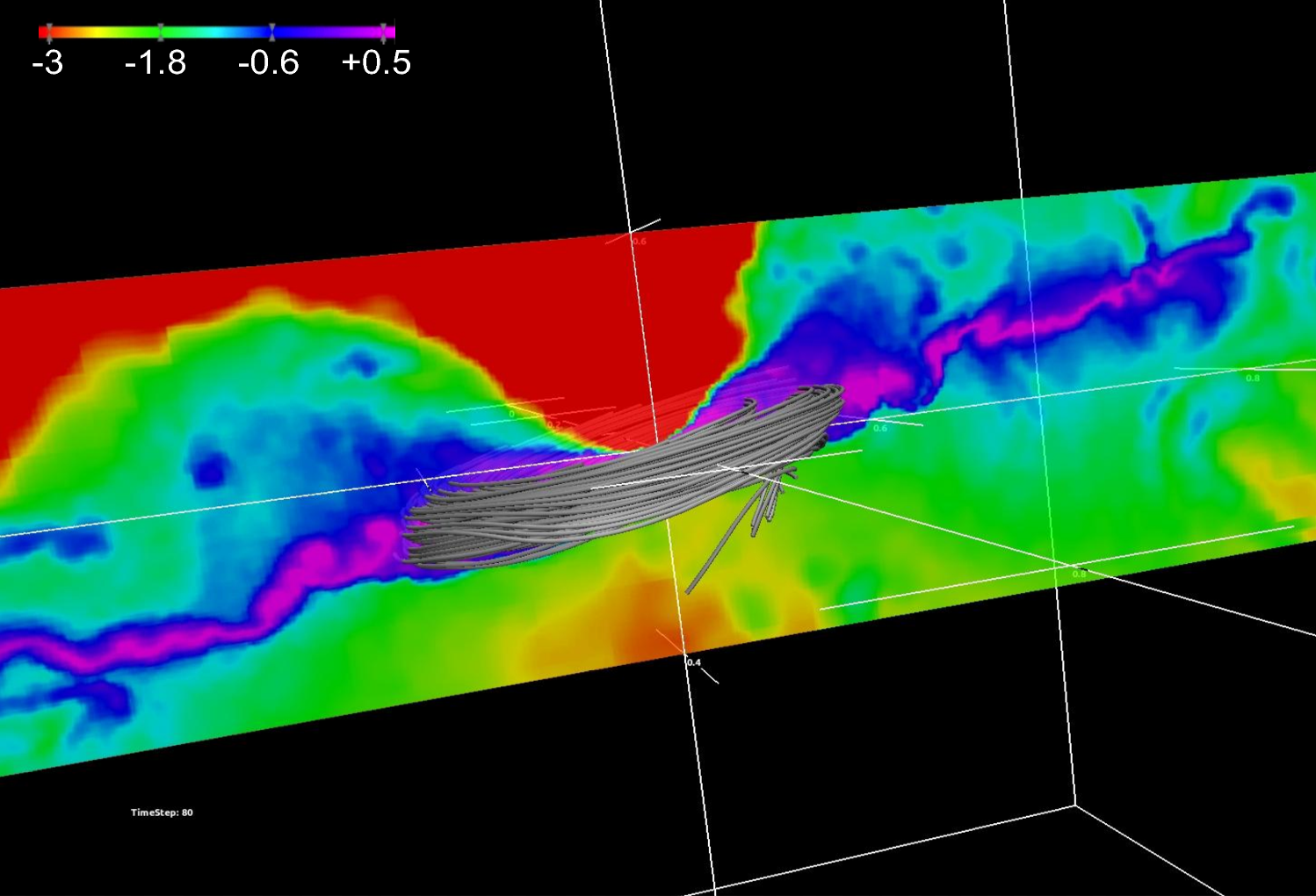} \end{center} \begin{center}
\includegraphics[scale=0.16]{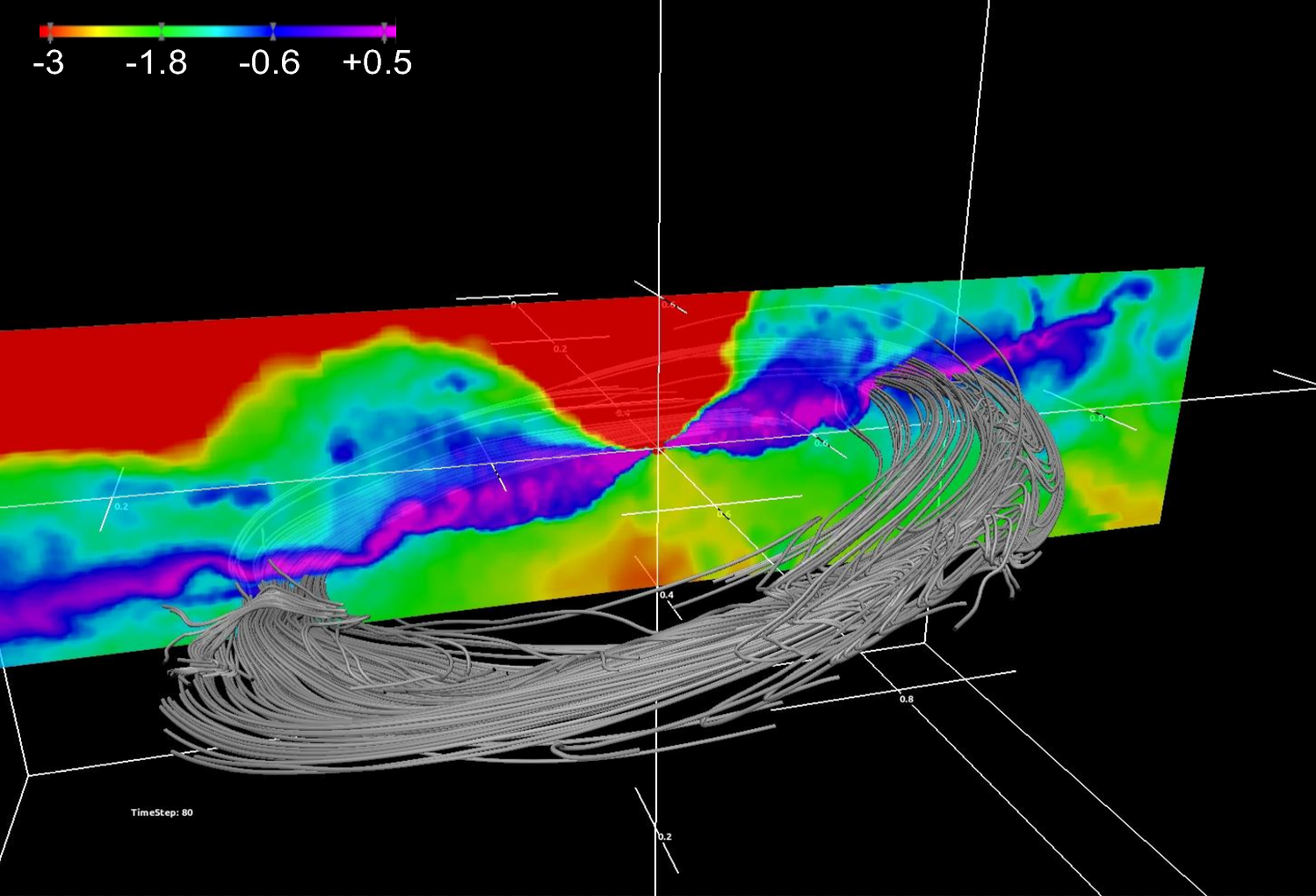}
\includegraphics[scale=0.16]{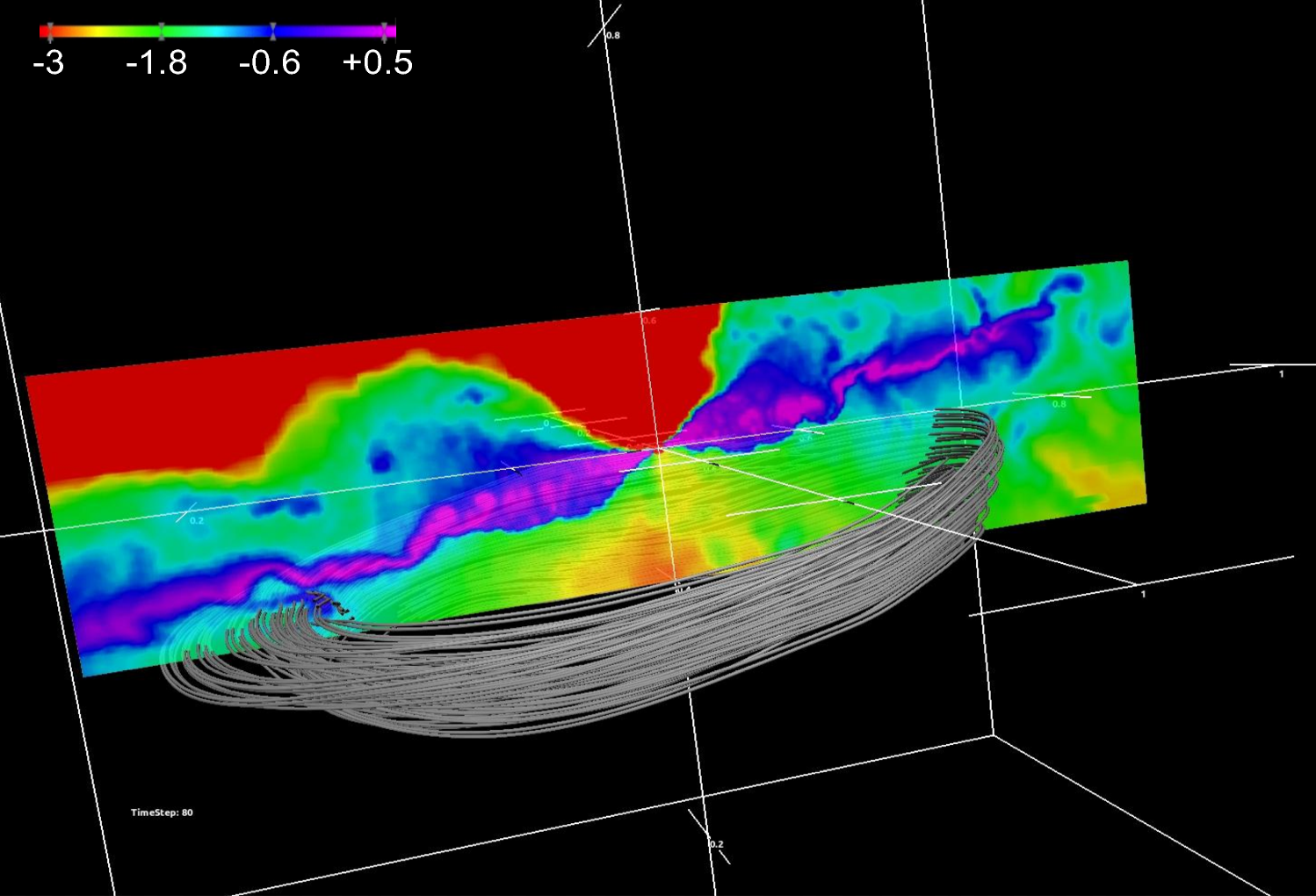} \end{center} \caption{This
figure shows magnetic field geometry in different regions of the disk. The
density is shown in the $XZ$ plane, with values given in
$\log(\rho/(4.1\cdot10^6{\rm cm}^3/m_u)$. The top two panels show magnetic
field geometry in the central region of the disk ($R \sim 0.1\,$pc), whereas
two bottom panel focus on outer regions of the disk ($R \sim 0.5\,$pc). While
magnetic field geometry is mostly uniform, the gas density show irregular
structure.} \label{fig:strong_bfield} \end{figure*}

\begin{table} \caption{This table shows geometry of initial conditions: Column
1 (ID) is the name of a run, Column 2 ($R_{\rm cl})$ show adius of a cloud,
Column ($v_x$) is cloud infall speed in km/s, and finally last coumn ($b$) is
the cloud's impact parameter.} \label{tbl:IC} \begin{center}
\begin{tabular}{cccc} \hline ID & $R_{\rm cl}$ [${\rm pc}]$& $v_x$ [${\rm
km/s}$] & $b$ [${\rm pc}$] \\ \hline\hline C01 & 3.5 &  120 & 2 \\ V01 & 3.5 &
30 &  3 \\ V02 & 3.5 & 50 & 3 \\ \hline \end{tabular} \end{center} \end{table}
We conduct three high resolution simulations with $1.6\times10^7$ particles in
which a magnetized gas cloud with mass and radius $3.5\,$pc and
$8.8\times10^4\,{\rm{M}}_\odot$, respectively, is collided with a $3.5\times
10^6\,{\rm{M}}_\odot\,$SMBH. The geometry of the initial conditions is taken
from \citet{2011MNRAS.412..469A} and repeated in Table \ref{tbl:IC}; in
particular, we use initial conditions from their simulations C01, V01 and V02,
where a molecular cloud has an impact parameter of $2\,$pc, $3\,$pc and $3\,$pc
respectively.  On top of these initial conditions we also impose a uniform
magnetic field that threads both the cloud and vacuum regions. The magnetic
field strength is such that the resulting magnetization inside the cloud is
$\beta = 1$, which corresponds to $|B|\approx100\,\mu$G and dimensionless
mass-to-flux ratio  $\zeta\approx5$, where
$\zeta=(M/\Phi)/\sqrt{5/(9\pi^2G)}\,$\citep{1976ApJ...210..326M,2004RvMP...76..125M}.
This field strength corresponds to the large-scale field in the Galactic Centre
\citep{1987AJ.....94.1178Y,1989ApJ...343..703M,2010Natur.463...65C}. The
initial magnetic field orientation was such that each of the components of
magnetic field have the same magnitude, namely $B_x = B_y = B_z = B/\sqrt{3}$.
The vacuum is modelled with fluid $10^6$ times less dense than the cloud
density ($n=0.01\,{\rm{cm}}^{-3}$), which we also use it as a floor density to
avoid local density contrasts larger than $\sim 10^7$ that our code cannot deal
with due to use of single precision floating point arithmetics.

The computational domain is a periodic box with $32\times32\times32\,$pc$^3$
volume, which is large enough not to influence physical processes occurring in
sub-parsec regions. The mass and distance units were $[M]=10^5\,{\rm{M}}_\odot$
and $[R]=1\,$pc respectively, which sets the time unit
$[T]\approx0.047\,{\rm{Myr}}$, magnetic field units $[B]\approx5.40\,$mG, and
the speed unit $[V]\approx20\,{\rm{km/s}}$. The simulation lasted till
$T_{\rm{end}}=5.0$, which corresponds to $0.24\,$Myr or $4.7$ orbital periods
at $R=1\,$pc. The inner boundary conditions are applied only within $0.02\,$pc
from the SMBH, which we regard as the inner disk boundary, as follows. Any
particle within $0.01\,$pc is removed from the computational domain, and in the
transition region between $0.01\,$pc and $0.02\,$pc we set the density to the
floor value, and both the velocity and the magnetic field to zero.

\section{Results}\label{sect:results}

\subsection{Geometry of the collision}

A collision between a molecular cloud and a supermassive black hole is a
violent event occurring on dynamical time-scale. Since fluid elements generally
have non-zero angular momentum, the natural outcome of such an event is a
formation of a disk.  Hydrodynamical simulations of such collision event
robustly show a formation of an eccentric disk around SMBH with the disk
geometry being dependent on the initial conditions \citep{2011MNRAS.412..469A}.
Our aim in this work is to study similar event but in strongly magnetised
regime in which the initial magnetic pressure in the cloud is in equipartition
with the gas thermal pressure.

In Fig.\,\ref{fig:V01dens} we show snapshots of the gas density in the $XY$
plane at $t$=0, 47, 94, and 240 thousand years. In the first hundred thousand
years, the cloud experience a violent collision with the black hole. In
particular, the bow-shock, which can be seen in the bottom-left panel as a
large curved region with density jump just above the disk, is formed by
isothermal shock guards the newly formed inner disk from the destructive effect
of the incoming fluid.  The outcome of this collision event is a formation of a
parsec-size gas disk with irregular density structure. Similar disks where
formed in other simulations as can be seen in the top two panels of
Fig.\,\ref{fig:V02C01dens}. This can be contrasted with
\cite{2011MNRAS.412..469A} where simulations C01 and V01 have final differently
shaped gas disks. Finally, in the bottom right panel of
Fig.\,\ref{fig:V02C01dens} we show the divergence error, $|\del\cdot{\bf
B}|h/|{\bf B}$ where $h$ is the cell size, in the final snapshot of V01
simulation.

%In Fig.\,\ref{fig:betab_map_V01dens}, similarly to the gas density, we show
%snapshots of gas magnetisation, $\beta^{-1}$, in the $XY$ plane at $t$=24, 47,
%94 and 240 thousand years. Initially the gas cloud is surrounded by strongly
%magnetised low density plasma. However, strong isothermal shocks that formed
%during the collision process combined with shear result in tenfold increase in
%the magnetisation of the incoming fluid. In particular, one can see strongly
%magnetised post-shock plasma inside the bow-shock region in the bottom-left
%panel of Fig.\,\ref{fig:betab_map_V01dens}. The inner disk, which has been
%formed by this time, already shows some signs of dynamical evolution, which is
%visible via azimuthally extended regions with lower magnetisation.

In Fig.\ref{fig:strong_bfield} we show  magnetic field geometry in different
regions of the disk at the end of the V01 simulation, which corresponds to
approximately 150 orbital periods at $R=0.1\,$pc\footnote{Since the inner
region of the disk is formed at approximately $t\sim 0.05$ million years, the
actual number of disk revolutions at $R\sim 0.1\,$pc is $\lesssim$100.}.  The
top-left panel show magnetic field lines originated in the central region of
the disk and extend above and below mid-plane. The magnetic field in this
regions is dominated by poloidal components. The top right panel shows
mid-plane magnetic field structure in the central region ($R\sim 0.1$pc).  The
field lines appear regular and tightly winding in azimuthal direction, which is
result of strong Keplerian shear inside the disk. In the bottom-left panel, we
also show magnetic field in the mid-plane region but further away from the
center ($R\sim 0.5\,$pc).  While magnetic field is still stretched in azimuthal
direction, in contrast to the central regions it shows less regular structure.
In the bottom-right panel we show magnetic field in the disk corona, where
magnetic field shows regular large-scale azimuthal pattern.

%%%%%%%%%%%%%%%%%%%%%%%%%%%%%%%
\subsection{Vertical structure}\label{sect:vertical}
%%%%%%%%%%%%%%%%%%%%%%%%%%%%%%%

\begin{figure*} \begin{center}
\includegraphics[scale=0.32]{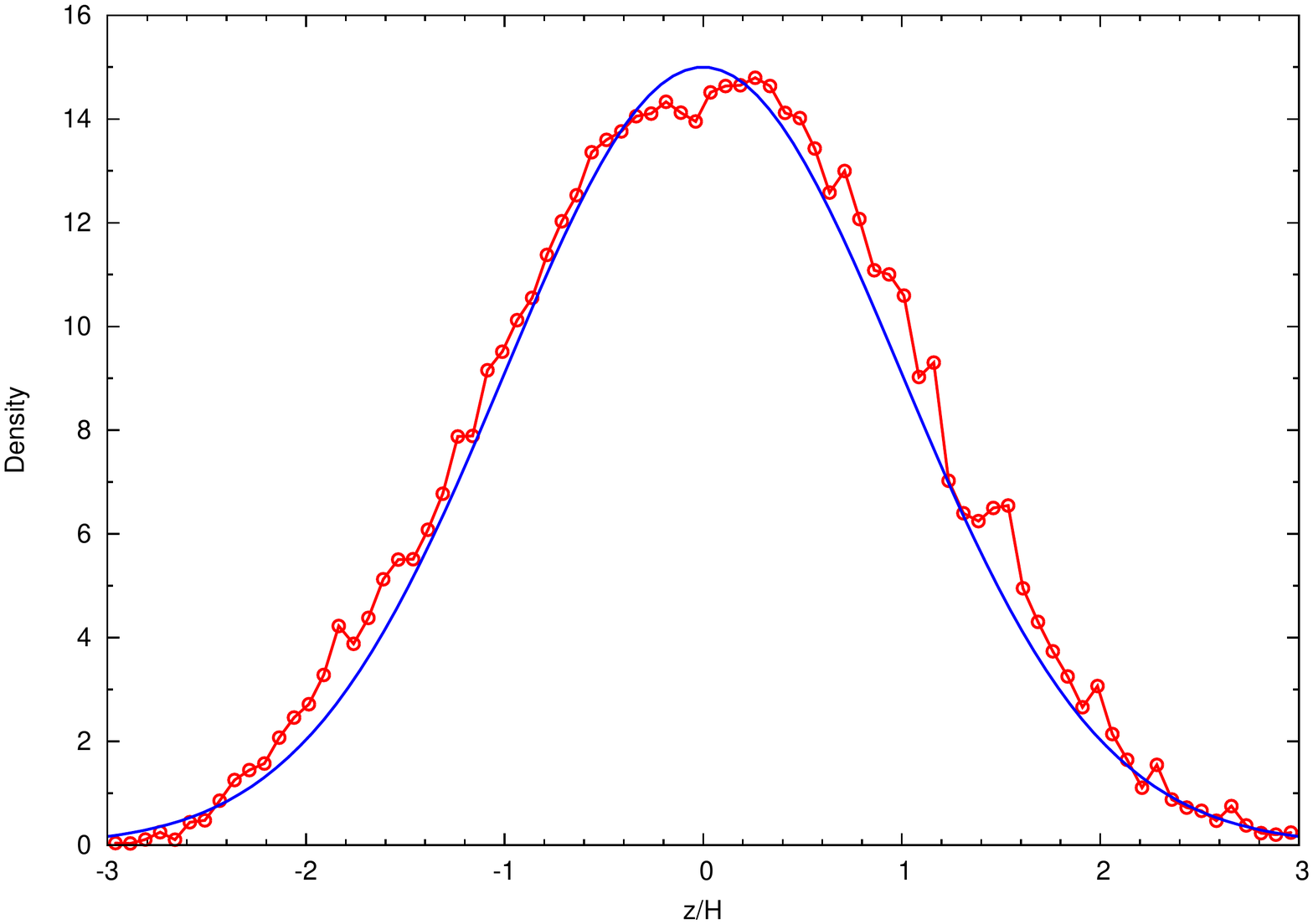}
\includegraphics[scale=0.32]{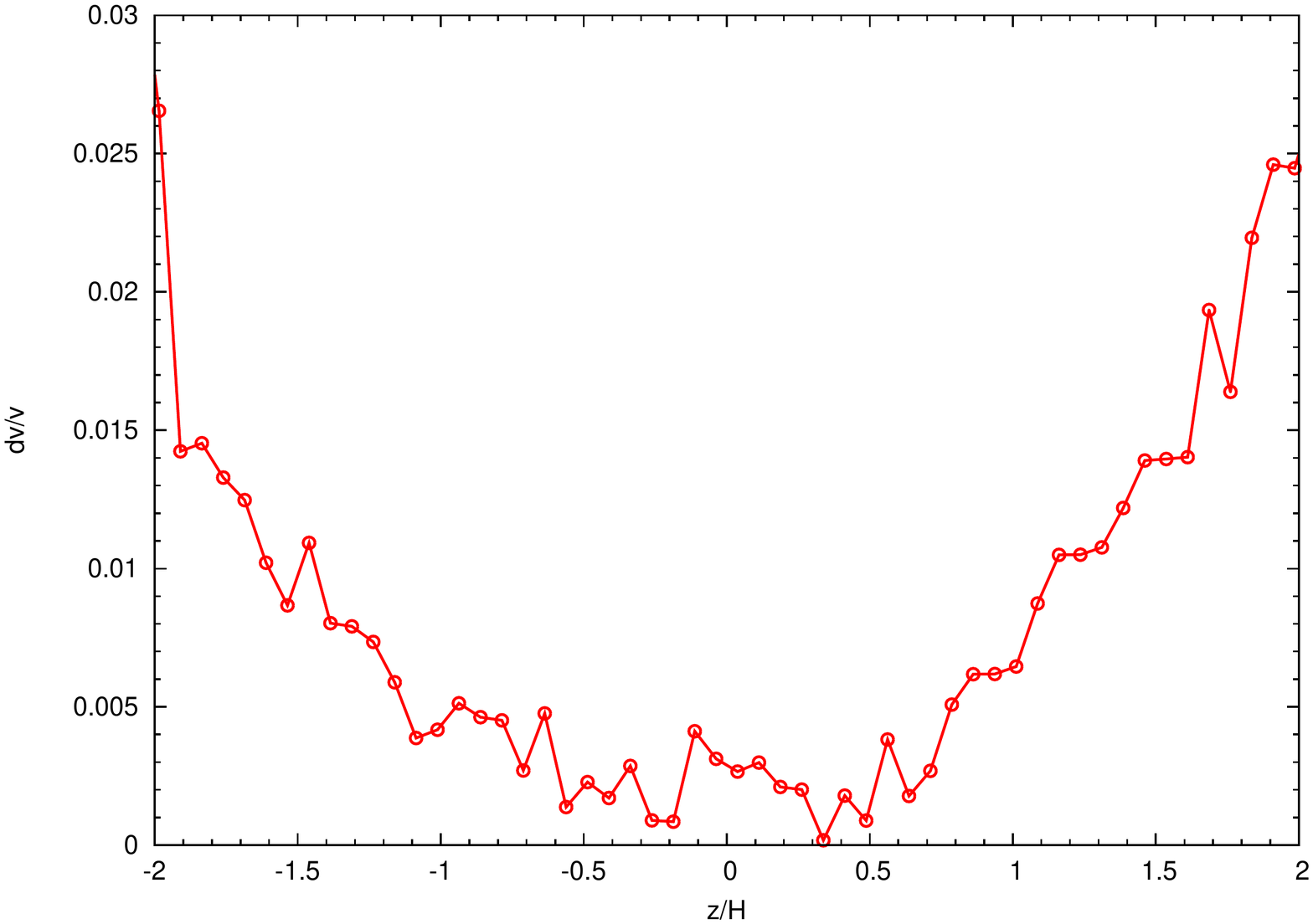}
\includegraphics[scale=0.32]{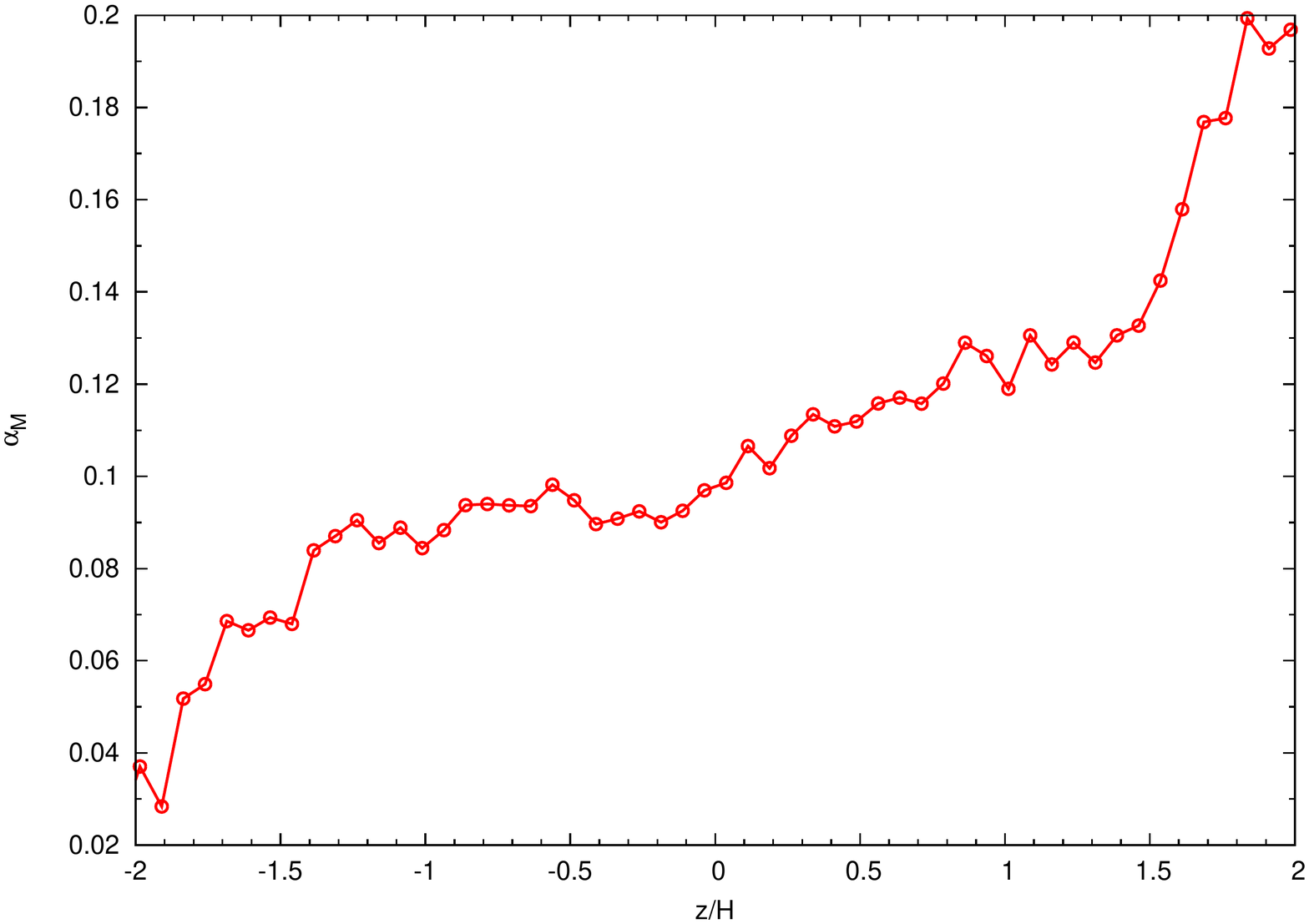}
\includegraphics[scale=0.32]{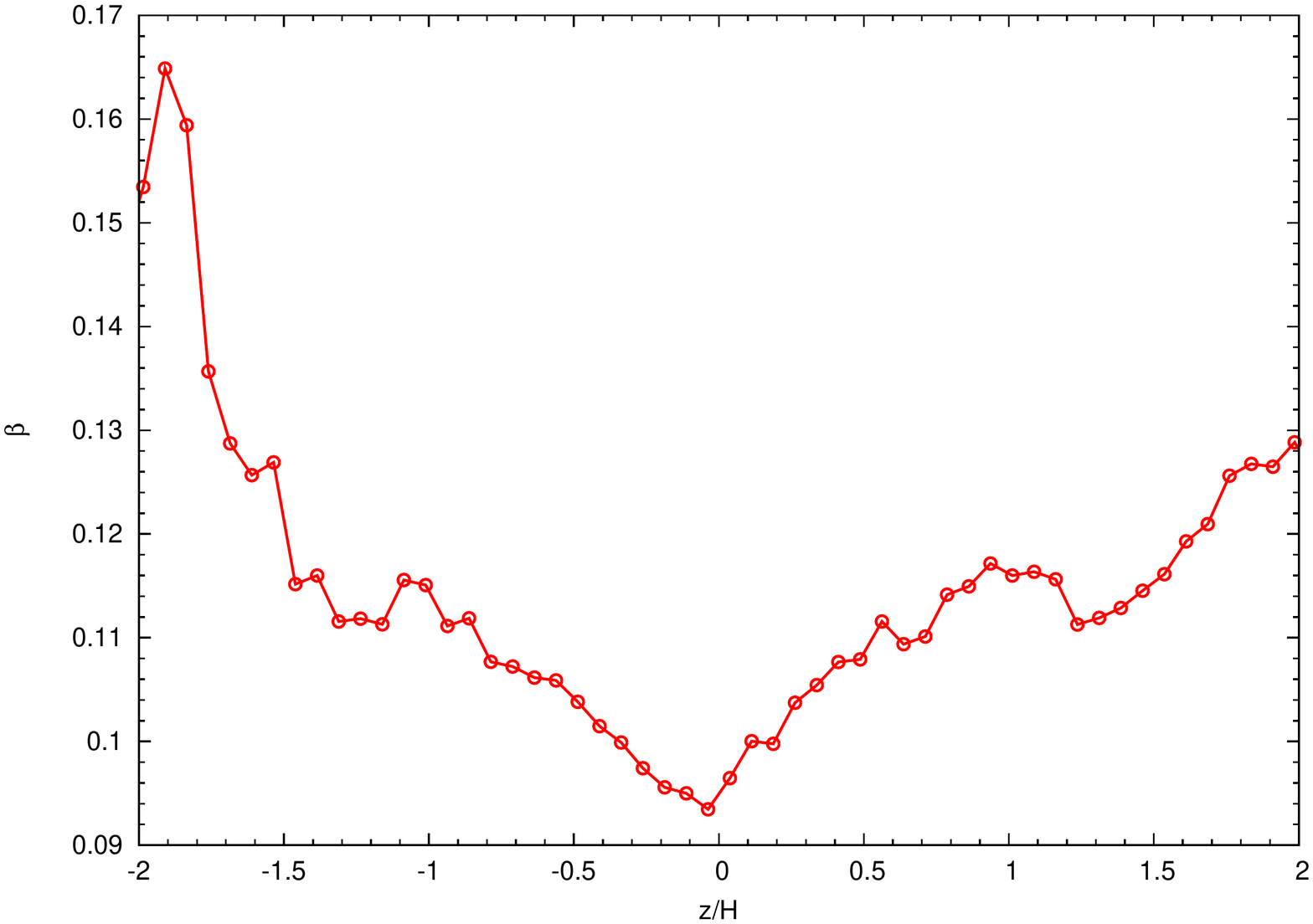} \end{center} \caption{This
figure shows vertical structure of the disk model at the end of the best
resolve simulation (V01). The top left panel shows density (red line with open
circle) and the fit of the expected density structure (blue line).  The top
right panel shows the deviation of the azimuthal velocity from the Kerplerian
value, the bottom left panel shows Maxwell stress, and the bottom right panel
show vertical dependence of the ratio of the gas pressure to the magnetic
pressure.} \label{fig:hydroVert} \end{figure*}

\begin{figure} \begin{center} \includegraphics[scale=0.37]{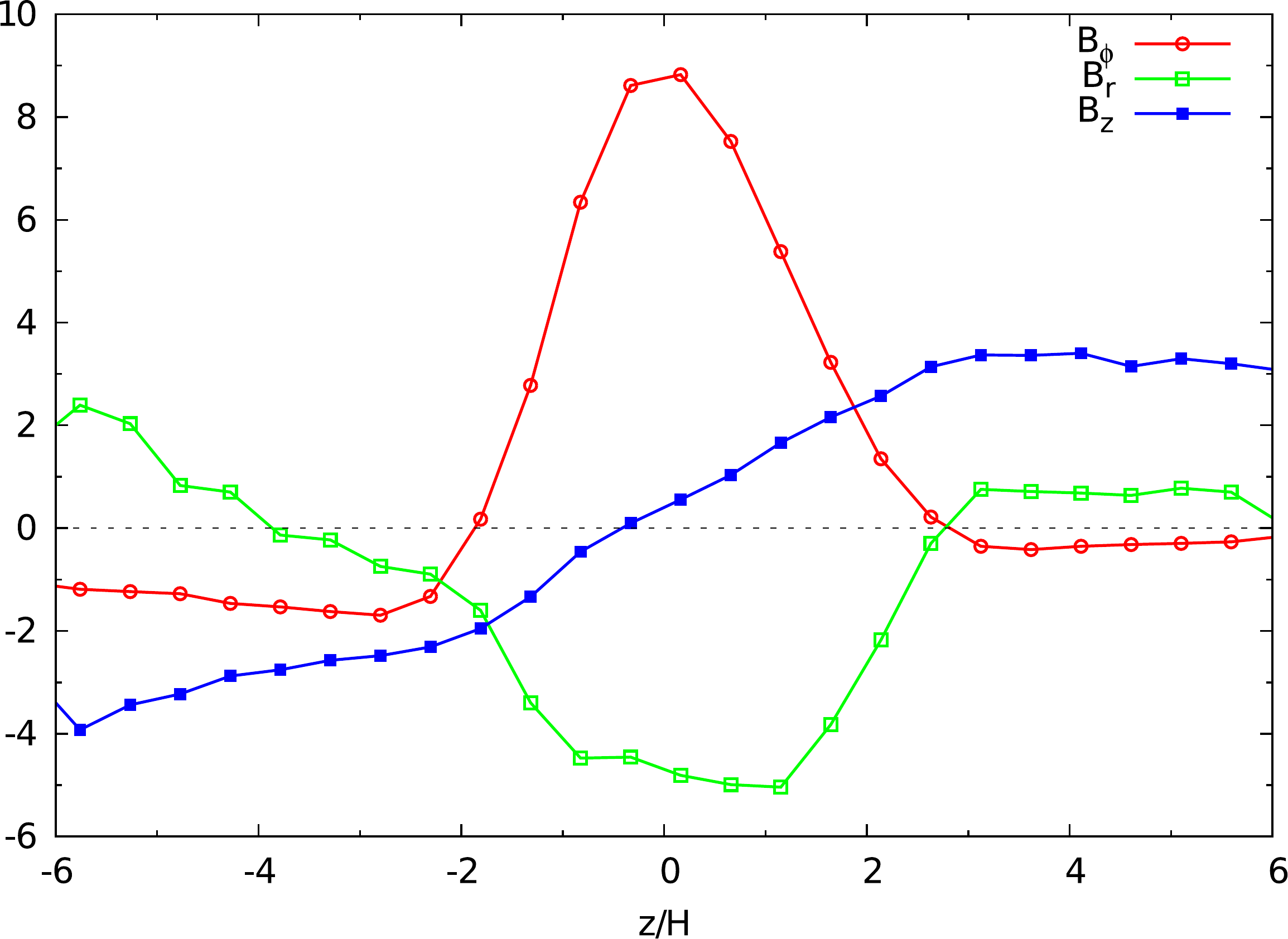}
\end{center} \caption{Vertical dependence of azimuthal (red lines with open
circles), radial (blue lines with open squared) and vertical (green line with
filled squares) magnetic field components. The dashed and solid lines show low-
and high-resolution simulations respectively.  Both radial and vertical
magnetic field components are $10\times$ magnified.} \label{fig:Bfield_H}
\end{figure}

%We would like to stress that while other simulations do not have enough
%resolution to accurately model axisymmetric MRI, this only affects a low term
%evolution of the disk. However, even our best resolved case, V01, has much
%smaller evolution time (less than 100 of orbits in the regions of interest) to
%really allow any long term effects to show up.

This section focuses on the vertical disk structure in our simulations.  In
particular, we studied vertical structure of the disk at $R\approx0.1\,$pc,
which is far enough from the inner boundary and close enough that the disk
performed approximately 100 orbital periods by the end of the simulation. One
of the crucial properties in magnetised disk simulations is the MRI quality
factor, $Q$, which is related to the number of resolution points, e.g.
grid-points, particles or mesh-cells, per MRI fastest growing mode wave-length.
With this number being too low ($Q \lesssim 8)$, the simulation may fail to
faithfully model magneto-rotational instability (e.g.
\cite{2011ApJ...738...84H}). Due to the nature of our simulations, it was
impossible \emph{a priori} to identify which of our simulations can faithfully
model long-term disk evolution. As a result, we computed MRI quality factors in
vertical and azimuthal directions at the end of our simulations, and check
which of the simulations were able to resolve MRI. Namely, we compute $Q_z =
\lambda_z/h$ and $Q_\phi = \lambda_\phi/h$, where $h$ is the size of a
resolution element and $\lambda_{z,\phi} \approx 2\pi
|B_{z,\phi}|/(\sqrt{\rho}\Omega)$.  \begin{table} \caption{MRI quality factor
for each of our simulations} \label{tbl:Qfac} \begin{center}
\begin{tabular}{ccc} \hline ID & $ Q_z$ & $Q_\phi$ \\ \hline\hline C01 & 3  &
26  \\ V01 & 11 & 60  \\ V02 & 6  & 48  \\ \hline \end{tabular} \end{center}
\end{table}

In Table\,\ref{tbl:Qfac} we show vertically averaged quality factors at $R
\approx 0.1\,$pc. This table demonstrates that all simulations have $Q_\phi
\gtrsim 8$, which means they can faithfully model non-axisymmetric MRI.
However, only V01 simulation qualifies when it comes to axisymmetric MRI, and
therefore we focus our study of vertical structure on this simulation. 

We compute scale-height, $H$, at this radius by fitting an isothermal density
profiles in approximately two scale-heights. The resulting scale-height is
$H\approx0.01$, which gives $H/R\approx0.1$ (top-left panel in
Fig.\,\ref{fig:hydroVert}). The radial temperature dependence is expected to
produce disks with the scale height $H/R=0.03\,\sqrt{1+\beta^{-1}}$ which, for
$\beta^{-1}=P_{\rm m}/P_{\rm g}\approx10$ found at $R=0.1\,$pc, gives

We studied vertical structure of the disk in V01 simulation at
$R\approx0.1\,$pc, which is far enough from the inner boundary and close enough
that the disk performed approximately 100 orbital periods by the end of the
simulation. We compute scale-height, $H$, at this radius by fitting an
isothermal density profile in approximately two scale-heights. The resulting
scale-height is $H\approx0.01$, which gives $H/R\approx0.1$ (top-left panel in
Fig.\,\ref{fig:hydroVert}). The radial temperature dependence is expected to
produce disks with the scale height $H/R=0.03\,\sqrt{1+\beta^{-1}}$ which, for
$\beta^{-1}=P_{\rm m}/P_{\rm g}\approx10$ found at $R=0.1\,$pc, gives
$H/R\approx0.1$, consistent with the simulation data (bottom-right panel in
Fig.\,\ref{fig:hydroVert}). We also studied the deviation of azimuthal velocity
form the Keplerian velocity at $R=0.1\,$pc as a function of height, and found
that azimuthal velocity variations are less than a percent for $|z| < H$
(top-right panel in Fig.\,\ref{fig:hydroVert}). It is therefore justifiable to
assume that the disk angular velocity is constant on cylinders. Finally, in the
bottom left panel of the Fig.\,\ref{fig:hydroVert} we show Maxwell stress
$\alpha_M = -\langle B_r B_\phi \rangle/P_{\rm tot}$  which is approximately
$0.1$ within the scale-height.
%eg:add

%In case of magnetic pressure dominated disk, the magnetic tilt angle,
%$\theta_M$, can be expressed in term of $\alpha_M$. Indeed, for $\beta \ll 1$
%we can write $\alpha_M \approx -\langle B_r B_\phi \rangle/\langle B^2/2
%\rangle

%via $\theta_M = \arcsin(\alpha_M)/2$. Setting $\alpha_M = 0.1$, we obtain
%$\theta_M \approx 2.8$ degrees which shows that magnetic field is tightly
%winded along azimuthal direction.

In Fig.\,\ref{fig:Bfield_H} we show azimuthally averaged magnetic field. This
figure show that the magnetic field is confined within few scale-heights of the
mid-plane. The magnetic field is dominated by the azimuthal component that is
an order of magnitude larger than the radial one. Vertical component, $B_z$, is
much smaller compared to both $B_r$ and $B_\varphi$ for $|z|\lesssim\,H/2$ (in
this figure both $B_r$ and $B_z$ strength are magnified by a factor of 10).

All of our simulations show similar vertical confinement of the field, which
can be interpreted as a result of the disk formation: a combination of
isothermal shocks that amplify magnetic field and Keplerian shear which
generates strong azimuthal field component. However, we would like to stress
that the field confinement in our best resolved model (V01) is in a good
agreement with \cite{2008A&A...490..501J} -- hereafter referred to as JL, who
find similar results in their shearing box models. In the JL shearing-box
simulations, which were performed with a grid-based Pencil Code, the disk field
was initially in equipartition with the gas pressure, but evolved by Parker and
magnetorotational instabilities to a magnetic field configuration in the
vertical direction similar to what we see in our disk which is formed via a
collision of a magnetized gas cloud with the black hole. It is significant that
the two simulations that are so different in their approach give vertical
structure of $B_\varphi$ that is in a good agreement with each other. In
particular, the azimuthal component changes sign at 2-3 scale-heights above the
mid-plane\footnote{In Fig.\,7 of JL the scale-height can be increased by
$\sqrt{2}$ due to extra support provided by the magnetic pressure.}, and the
strength of the mid-plane field is ten times the value of the reversed field. 

%Whether the field confinement lasts for many hundred of orbits remains to be
%studied via either shearing-box or isolated global-disk models.

%If we were to take that the resulting structure is not only the result of the
%collision but also the outcome of non-linear MHD evolution, we can draw
%following The strength of $B_z$ remains much smaller compared to both $B_r$
%and $B_\varphi$ for $|z|\lesssim\,H/2$. Furthermore, $B_z$ mostly consists of
%small-scale fluctuation and $\langle{B_z^2}\rangle\gg\langle{B_z}\rangle^2$,
%with the mean value not showing convergence in low-resolution simulations. 

%The vertical structure of the magnetic field can be used to explain the global
%magnetic field geometry in Fig.\,\ref{fig:strong_bfield}: in the disk
%mid-plane the magnetic field lines tightly wind around azimuthal direction
%many times before migrating first radially inwards and then above mid-plane.
%Inspection of $B_z$ shows that a field line beginning just above/below the
%mid-plane, will leave the disk also above/below the mid-plane.

\begin{figure}[] \begin{center} \includegraphics[scale=0.19]{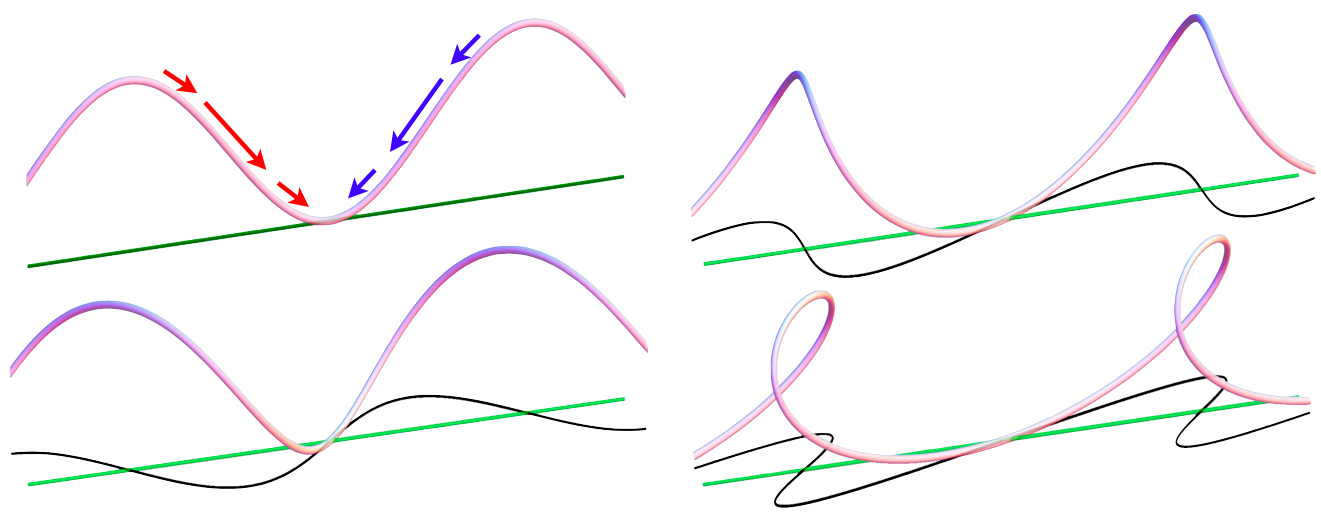}
\end{center} \caption{Sketch of the evolution of a magnetic field line subject
to Parker instability, Coriolis force and Keplerian shear. The green straight
tube in all panels shows the original field line along $\varphi$ direction in
the mid-plane, the white-blue 3D tube depicts the actual field line, and its
projection onto the $r-\varphi$ plane is shown in black. The top-left panel
shows the undulant distortion of the original field line due to Parker
instability in the $\varphi-z$ plane. The fluid elements that slide along the
field lines (red and blue arrows) towards the mid-plane are acted upon by the
Coriolis force transforming the field line into a helical form.  Projection of
the field line onto $r-\varphi$ plane shows creation of a radial component
(bottom-left). Stretching of the radial component by Keplerian shear
regenerates the azimuthal component (top-right). Finally, subsequent stretching
of the line generates oppositely oriented magnetic field above mid-plane and
increases the strength of the azimuthal component of magnetic field in the
mid-plane (bottom-right).} \label{fig:parker} \end{figure} It is likely that
the process responsible for flux confinement is of a similar nature as
described in JL. We sketch it in Fig.\,\ref{fig:parker}. The strong azimuthal
magnetic field is subject to Parker instability. As the fluid elements slide
along the field line towards the mid-plane (top-left) they are acted upon by
the Coriolis force \citep{2002A&A...386..347H}, and due to this the line
becomes helical (bottom-left).  This generates radial field, which via
Keplerian shear regenerates the azimuthal field that was lost to Parker
instability (top-right). Further shearing generates oppositely oriented
azimuthal field above the mid-plane which reconnects with the field of the
original orientation.  This decreases its magnitude or even reverses its
direction, whereas the field at the mid-plane is amplified while maintaining
its original direction (bottom-right).

%%%%%%%%%%%%%%%%%%%%%%%%%%%%%%%
\subsection{Radial structure}\label{sect:radial}
%%%%%%%%%%%%%%%%%%%%%%%%%%%%%%%
The radial evolution of the disk is driven by the flow of matter from the outer
to the inner regions. This flow is enabled by the effective viscosity from
magnetohydrodynamical stresses. Within the viscous time-scale, the disk reaches
steady radial structure which is computed by applying conservation laws and
theory of steady thin disks \citep[e.g.][]{2002apa..book.....F}. {Here we
present some numerical evidence for an extra constraint, the conservation of
azimuthal magnetic flux, $\Phi=\int\,B_\varphi\,dS$ where
$dS\,=dz\,v_{\rm{r}}\,dt$. The set of equations that describe disk's radial
structure is \begin{eqnarray} \label{eqn:massdot} \dot
M&=&2\pi\,R\,\Sigma\,v_{\rm{r}},\\ \label{eqn:fluxdot} \dot
\Phi&=&2H\,B_\varphi\,v_{\rm{r}}, \\ \label{eqn:maccdot} \dot
M&=&3\pi\,\alpha_{\rm{acc}}\,\Sigma\,H^2\,\Omega.  \end{eqnarray} Here,
Eq.\,(\ref{eqn:fluxdot}) describes the frozen-in condition of magnetic
field\footnote{The condition can also be derived from the induction equation by
consideration of the radial advective flux of the vertically integrated
azimuthal magnetic field, $\int\,B_\varphi\,dz$.}. In what follows we assume
that the right hand side of these equations are constants for steady-state
disks. We also assume that accretion viscosity $\alpha_{\rm acc}$ is set by
Maxwell stresses \begin{equation}
\alpha_{\rm{acc}}=\alpha_{\rm{m}}=-\frac{\langle
B_r\,B_\varphi\rangle}{4\pi\,P}, \label{eqn:alpha_acc} \end{equation} where
$P=\Omega^2\,H\,\Sigma$ is the total pressure. The disk scale-height is
\begin{equation} H=H_0\sqrt{1+\beta^{-1}}, \label{eqn:scaleH} \end{equation}
where $H_0=c_{\rm{s}}/\Omega$ is a hydrodynamical scale-height.  According to
the numerical results from the previous section, the magnetic pressure is
entirely dominated by $B_\varphi$, which gives
$P_{\rm{m}}\approx\,B_\varphi^2/8\pi$.  \begin{figure} \begin{center}
\includegraphics[scale=0.33]{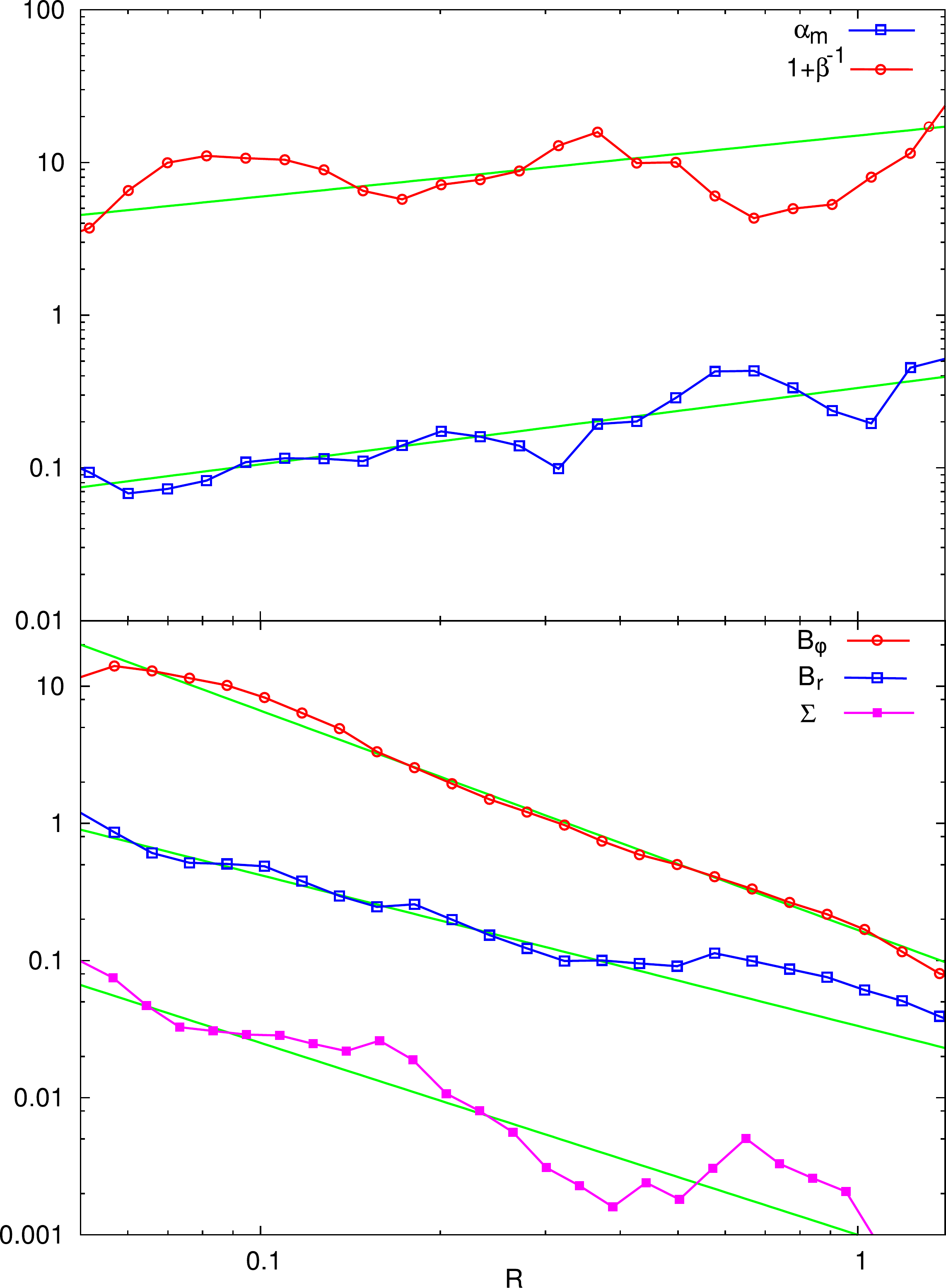} \end{center} \caption{Radial
structure of strongly magnetized disks in our simulations.  The upper panel
displays disk magnetization, (red line with open circles) and dimensionless
viscosity generated by Maxwell stresses (blue line with open squares). The
lower panel shows radial dependence of surface density (magenta line with
filled squared), and mid-plane $B_\varphi$ (red line with open circles) and
$B_r$ (blue line with open squares); $B_z$ has the same scaling and magnitude
as $B_r$ and is not shown here.  The green lines show expected radial
dependence from our analytical model.} \label{fig:alpha_beta_R} \end{figure}

Using these equations and noting that
$(1+\beta)\,B_\varphi^2/8\pi\,=\Omega^2\,H\,\Sigma$, we derive the radial
dependence of disk scale-height \begin{equation}
\frac{H}{R}\,=\,\left(\frac{3\pi^2\,G\,\dot{M}\,(1+\beta)}{40\,\alpha_{\rm{m}}\,\zeta^2\,(\Omega\,R)^3}\right)^{\frac{1}{5}},
\label{eqn:scale_height} \end{equation} where we write
$\dot\Phi=\sqrt{5/(9\pi^2G)}\dot{M}/\zeta$. In the limit, $\beta\ll1$ the
radial dependence of disk magnetization is \begin{equation}
\beta^{-1}\,=\,\left(\frac{3\pi^2\,G\,\dot{M}}{40\,\alpha_{\rm{m}}\,\zeta^2(\Omega\,R)^3}\right)^{\frac{2}{5}}
\left(\frac{R}{H_0}\right)^2.  \label{eq:magnetization} \end{equation} In a
general case, $H_0$ is self-consistently computed by solving radiative transfer
equation in the vertical direction. Therefore, the magnetization depends on the
thermal properties of the disk. However, in our simulations $H_0\,=\,0.03\,R$
from which we have $\beta^{-1}\propto\,R^{3/5}\,\alpha_{\rm{m}}^{-2/5}$.

Since the radial dependence of $\alpha_{\rm{m}}$ is not possible to establish
from the first principles, we obtain it empirically. In our numerical
experiments $\alpha_{\rm{m}}\propto\,R^{1/2}$ is consistent with the data. This
relationship predicts that the disk magnetization should decrease with radius,
$\beta^{-1}\propto\,R^{2/5}$, in agreement with our simulations (upper panel in
Fig.\,\ref{fig:alpha_beta_R}).  This radial dependence of $\alpha_{\rm{m}}$ is
specific to our simulations which we use for consistency check. However, a
similar dependence was found by \cite{2011ApJ...735..122F} in the case of
weakly-magnetised disks. In a realistic disk, however, we expect that
$\alpha_{\rm{m}}$ is a function of the local $\beta$, which will be subject of
subsequent research.

Using equations above, we derive the radial dependence of $B_\varphi$, $B_r$
and $\Sigma$. To derive $B_r$, we use the fact that in our disks the Maxwell
stresses in Eq.\,(\ref{eqn:alpha_acc}) are dominated by the mean field,
$\langle{B_rB_\varphi}\rangle\approx\langle{B_r}\rangle\langle{B_\varphi}\rangle$.
The resulted radial dependences are \begin{eqnarray} \label{eqn:SigmaR}
\Sigma&=&\left(\frac{1600}{2187\pi^9}\frac{\dot{M}^3\zeta^4}{G^2(1+\beta)^2}\right)^{\frac{1}{5}}\frac{(\Omega\,R)^{\frac{1}{5}}}{\alpha_{\rm{m}}^{3/5}R}\propto\,R^{-\frac{7}{5}},\\
\label{eqn:BphR}
B_\varphi&=&\left(\frac{512\sqrt{5}}{27\pi}\frac{\dot{M}^2\zeta}{\sqrt{G}(1+\beta)^3}\right)^{\frac{1}{5}}\frac{(\Omega\,R)^{\frac{4}{5}}}{\alpha_{\rm{m}}^{2/5}R}\propto\,R^{-\frac{8}{5}},\\
\label{eqn:Br}
B_r&=&\frac{\alpha_{\rm{m}}}{2}B_\varphi\propto\,R^{-\frac{11}{10}}.
\end{eqnarray} These equations are consistent with our simulations throughout
most parts of the disk (bottom panel in Fig.\,\ref{fig:alpha_beta_R}), except
in the regions close to the disk inner boundary. We also do not expect the model
to hold for $R\gtrsim0.4\,$pc, where the viscous time estimated using steady
thin disk approximation is much larger than the duration of simulation.
Agreement with the analytical model beyond this radius implies that disk
evolution there occurs at higher than viscous rate derived from the steady thin
disk theory. We also note that the gas density distribution in the disk is not
steady, but exhibits clumpy and filamentary structures (right panel in
Fig.\,\ref{fig:strong_bfield}). This is also reflected in the irregularity of
the surface density profile in Fig.\,\ref{fig:alpha_beta_R}. Nevertheless,
agreement of radial dependence between the model and simulations indicates that
azimuthal magnetic flux is conserved in MLADs during accretion.

\begin{table}[h] \caption{Accretion rate and effective viscosity}
\begin{center} \begin{tabular}{c c c c } \hline \hline $R\,$[pc] &
$\dot{M}\,$[M$_\odot$/yr] & $\alpha_{\rm{acc}}$ & $\alpha$ \\ \hline 0.05 &
0.035 & 0.06 & 0.03  \\ 0.1  &  0.066 & 0.09 & 0.14  \\ 0.2  &  0.070 & 0.22 &
0.18  \\ 0.4  &  0.085 & 0.37 & 0.31  \\ \hline \hline \end{tabular}
\end{center} \label{tbl:mdot} \end{table} To verify that the mass accretion is
physical, we extract azimuthally and vertically averaged $\Sigma$, $H$,
$\dot{M}$ at different radial locations and use Eq.\,\ref{eqn:maccdot} to
calculate $\alpha_{\rm{acc}}=\dot{M}/(3\pi\Sigma\,H^2\Omega)$.  If the
accretion is driven by magnetohydrodynamical stresses, this value should be
comparable to azimuthally and vertically averaged sum of Maxwell and Reynolds
stresses ($\alpha$). We show results in Tab.\,\ref{tbl:mdot}, which shows good
agreement between measured ($\alpha$) and derived viscosity ($\alpha_{\rm
acc}$) coefficients.  This reinforces our confidence that the mass accretion is
indeed driven by magnetohydrodynamical stresses. Finally, using data from this
table we estimate viscous timescale for a parsec size accretion disk to be
$t_{\rm visc}= (R/H)^2\alpha^{-1}\Omega^{-1}\approx 10^6\,$ years, where we use
$\alpha = 0.2$ and $H/R = 0.2$ at $R = 1$\,pc. 

\section{Fragmentation}\label{sect:fragmentation} 

The striking result that $\Sigma$ and $H$ do not depend on the thermal
properties of MLAD allows a robust estimate of its macroscopic gravitational
stability.\footnote{Some of the gas clumps and filaments may form stars even if
the disk is globally stable.} Its fragmentation boundary is determined by two
parameters: $\dot{M}$ and dimensionless mass-to-flux ratio $\zeta$. The latter
one allows to compute magnetic flux accretion rate from the mass accretion
rate, $\dot\Phi\propto\dot{M}/\zeta$, since the magnetic field is frozen in the
fluid.  Using Eq.\,(\ref{eqn:scale_height}) and Eq.\,(\ref{eqn:SigmaR}), the
Toomre-$Q$ parameter is \citep{1964ApJ...139.1217T,1965MNRAS.130...97G}
\begin{equation}
Q=\frac{\Omega^2\,H}{\pi\,G\,\Sigma}=\left[\frac{6561\pi^6}{64000}\frac{(1+\beta)^3}{G^2\,\dot{M}^2\,\zeta^6}\alpha_{\rm{m}}^2(\Omega\,R)^6\right]^{\frac{1}{5}}.
\label{eqn:toomre_Q} \end{equation} If we write $\dot{M}$ in terms of Eddington
luminosity, $l_{\rm{E}}=L/L_{\rm{Edd}}$, and radiative efficiency,
$\epsilon=L/(\dot{M}c^2)$, \begin{equation} \dot M=\frac{4\pi G
M}{\kappa_{\rm{es}}c}\frac{l_{\rm{E}}}{\epsilon}, \label{eqn:mdot_eqn}
\end{equation} where $\kappa_{\rm{es}}\approx0.4\,{\rm{c}m}^2/{\rm{g}}$ is
electron scattering opacity, and $c$ is speed of light, we obtain
\begin{equation}
Q=\left[\frac{6561\pi^4}{4^510^3}\frac{\kappa^2\,c^2(1+\beta)^3}{G^2}\frac{\alpha_{\rm{m}}^2}{\zeta^6}\left(\frac{\epsilon
}{l_{\rm{E}}}\right)^2\Omega^2\right]^{\frac{1}{5}}.  \label{eqn:toomre_Q1}
\end{equation}

Using Eq.\,(\ref{eqn:toomre_Q1}) we find the fragmentation boundary beyond
which $Q<1$, \begin{equation}
R_{\rm{frag}}\approx2.09\left(\frac{M_6\,\alpha_{0.1}^2\,\epsilon_{0.1}^2}{\zeta_{10}^6\,l_{\rm{E}}^2}\right)^\frac{1}{3}\,{\rm{pc}},
\label{eqn:Rfrag} \end{equation} where we used $\epsilon =
0.1\,\epsilon_{0.1}$, $\alpha_{\rm{m}}=0.1\,\alpha_{0.1}$,
$\zeta=10\,\zeta_{10}$ and $M=10^6\,M_6\,{\rm M}_\odot$. The radial dependence
of enclosed mass and $H/R$ within the fragmentation radius is given by
\begin{eqnarray} \label{eqn:enclosed_mass}
\frac{M_{\rm{disk}}}{M}&=&\frac{\sqrt{10}\pi\,r^{9/10}}{3\zeta}\approx0.331\,\frac{r^{9/10}}{\zeta_{10}},\\
\label{eqn:enclosed_HoR}
\frac{H}{R}&=&\frac{3\pi\,r^{3/10}}{2\sqrt{10}\zeta}\approx0.149\,\frac{r^{3/10}}{\zeta_{10}},
\end{eqnarray} where we define $r=R/R_{\rm{frag}}$. It is worth noticing, that
at the fragmentation boundary, $M_{\rm{disk}}$ and $H/R$ depend only on
mass-to-flux ratio. 

In future work we will use our MLAD solution to model observations of AGN
accretion disk. Here, we briefly consider a parsec-sized disk in NGC1068
\citep{2004Natur.429...47J} as an eaxmple. This Seyfert 2 galaxy hosts an
$\sim10^7\,{\rm{M}}_\odot$ SMBH with a disk extending to a distances of
$\sim1\,$pc. The observed upper bound for $H/R\sim0.6$ and the hydrogen column
density
$N_{\rm{H}}\sim10^{25}\,{\rm{cm}}^{-2}\,$\citep{2004A&A...414..155M,2010MNRAS.406L...6K},
and its luminosity is $\sim0.4\,L_{\rm{Edd}}$ \citep{1994ApJ...428..124P} .
Here, we assume that this SMBH accretes at Eddington rate
($l_{\rm{E}}\approx1$) with 10\% radiative efficiency
($\epsilon_{0.1}\approx1$); we set $\alpha_{0.1}=1$.  We find that by setting
$\zeta=3$, we are able to obtain values for disk thickness and column density
that are consistent with observations. For this parameters, the MLAD thickness
at the edge of such disk is $H/R\approx0.15$.  While this is lower than the
observed value, it is plausible that strong magnetic fields contribute toward
increasing disk thickness. Finally, using the enclosed disk mass at this
location and assuming that the disk consists purely of atomic hydrogen, we
estimate $N_{\rm{H}}\approx1.3\times10^{25}\,{\rm{cm}}^{-2}$ which is
consistent with the observational data. Furthermore, the fragmentation boundary
is located at $\approx50\,$pc, which indicates that a parsec-sized disk is
stable to clumping.  This MLAD model predicts that such disk should have
magnetic field strength of $\sim100\,$mG.

\section{Conclusions}\label{sect:discussion}

In this paper we produce from first principles dynamically stable models of
accretion disks in a state of magnetic levitation. We show that such disks are
the natural outcome of in-fall of a massive magnetized molecular cloud onto
supermassive black hole. Such magnetically-levitating accretion disks (MLADs)
enable large accretion rates due to the large scale-height and
$\alpha\gtrsim0.1$. In our simulations, the geometry and strength of the
large-scale magnetic field are stable for at least $0.24\,$Myr, corresponding
to several hundred orbits at the disk inner edge. With measured accretion rates
of $\approx 0.05\,{\rm M}_\odot$/yr, this is more than 10\% of the disk
lifetime, supporting the claim that such magnetic fields structure is possibly
long lived.  The viscous time-scale of such magnetically levitating disk is
estimated to be few million years. Interestingly, this feature may help solve a
theoretical problem that was recently identified by \cite{2011MNRAS.tmp.1736A}
with respect to the formation of the stellar disc in our Galactic Centre. These
authors show that if, as according to the currently accepted scenario
\citep{2003ApJ...590L..33L, 2007MNRAS.379...21N, 2008Sci...321.1060B}, the
stellar disc formed as a result of fragmentation of the massive gaseous
accretion disc several million years ago, then a substantial gaseous remnant of
the accretion disc should survive to the present epoch, due to the expected
long viscous time of the standard Shakura-Syunyaev thin discs. Such a remnant
is not observed.  On the other hand, the expected short lifetime of the MLAD,
may solve the problem of the missing remnant  gas disk.

A unique property of magnetically-levitating disks is that their surface
density and scale-height are independent of the disk's thermal structure.  This
is expected because thermal effects are superseded by magnetic properties in
determining disk structure.  Magnetic levitation allows the disk to withstand
its own self-gravity to large distances. A strong dependence of the
fragmentation radius on the mass-to-flux ratio of the parent cloud permits a
scenario in which a tidal disruption of a magnetized cloud forms a magnetized
gas ring. The inner parts spreads inwards on a time-scale determined by global
magnetic stresses which fuel fast accretion onto the central supermassive black
hole, while the outer part fragments into stars.

%eg:add
A proper understanding of the field confinement requires both local and
global analysis. In accordance with JL, the field confinement appears to be a
local phenomenon and its stability is likely to depend on the relative strength
of vertical and azimuthal magnetic field, which itself depends on both
kinematics and magnetisation of the infalling matter. However our simulations
show that there are non-local processes which generate a global coherent
magnetic field structure. Its topology and strength is very important for
accretion flows near black-hole horizons
\citep{2008ApJ...678.1180B,2011MNRAS.418L..79T,2012MNRAS.423L..55T}. Therefore
it is also important to understand the long-term evolution of the field
topology across several decades in the disc radius we believe that both local
and global simulations are essential to our understanding of MLADs. 

\section*{Acknowledgements} We thank Daniel Price for help with SPLASH
\citep{2007PASA...24..159P}, John Clyne for help with VAPOR
\citep[{\tt{http://www.vapor.ucar.edu}}]{VAPORref}, and Tsuyoshi Hamada for
using DEGIMA GPU-cluster. We also thank Richard Alexander, Andrei Gruzinov and
Andrei Beloborodov for discussions, and the anonymous referee for the
insightful comments that helped to improve the manuscript. This work is
supported by the NWO VIDI grant \#639.042.607 and by NASA through a Hubble
Fellowship grant HST-HF-51289.01-A from the Space Telescope Science Institute,
which is operated by the Association of Universities for Research in Astronomy,
Incorporated, under NASA contract NAS5-26555.

\appendix
%%% \section*{Appendix}
\section{Numerical method}

In this appendix we demonstrate the ability of our numerical scheme to model
MHD flows. Our numerical method combines a moving-mesh approach
\citep{1988CoPhC..48...39T,2010MNRAS.401..791S} with a weighted particle MHD
scheme \citep{2011MNRAS.414..129G}. At every time-step a Voronoi mesh is
re-built on the set of particles which is used to solve equations of ideal MHD
in the same way as in the weighted particle scheme. Similar approach has also
been attempted in TESS \citep{2011ApJSTESS} and AREPO
\citep{2011MNRAS.tmp.1536P} moving-mesh codes. In contrast to these two
approaches but similarly to \cite{2011MNRAS.414..129G}, we add a source term to
the induction equation that restores Galliean invariance of scheme in the case
of $\del\cdot{\rm B} \ne 0$. This proved to be crucial to stabilize the
numerical scheme in the presence of strongly magnetised super-alfvenic flows.
The numerical code is pubcly available.\footnote{ {\tt
http://github.com/egaburov/fvmhd3d}} Here, we present validation of our
numerical method by means of two test problems: propagation of circularly
polarized Alfven wave and the linear regime of magneto-rotational instability
in a cylindrical disk.

  \subsection{Circularly polarized Alfven wave} This problem was first
presented by \cite{2000JCoPh.161..605T} as an exact non-linear test problem for
ideal MHD.  Following \cite{2000JCoPh.161..605T} we set the following initial
conditions. We use a periodic three-dimensional computational domain with the
total number of particles equal to $N_{\rm tot} = N_x \times  N_x/2 \times 16$,
where $Nx = 16, 32, 64$ and $128$. Particles were initially randomly sampled
from a uniform distribution and regularized with the Lloyd's algorithm (e.g.
\cite{2010MNRAS.401..791S}). The initial conditions are $\rho = 1$, $P_{\rm
gas} = 0.1$, $B_x = 1$, $v_x = 1$, $B_y = v_y = 0.1\sin(4\pi x)$, $B_z = v_z =
0.1\cos(4\pi x)$, which fits two wave-length into the $x$-direction. With these
initial conditions, the wave-length is resolved with an average of $6, 13, 26$
and $52$ mesh-points from the lowest to the highest resolution respectively.
The apparent discrepancy from the expected resolutions of $8, 16, 32$ and $64$
mesh-cells per wave-length is due to the initial particle distribution is not
being a simple cubic lattice, but rather a random distribution which was
relaxed by the Lloyd's algorithm. This relaxed distribution consists of
mesh-cells which can be approximated by regular convex polyhedra with large
number of faces ($\gtrsim 15$). The effective size of such mesh-cell can be
approximated by the diameter of a sphere having the same volume as the cell
itself, and this in turn increases the effective size of the mesh-cell by
approximately $\sqrt[3]{6/\pi} \approx 1.24$ compared to a simple cubic cell,
while keeping the total volume the same.

  \begin{figure*}[] \begin{center}
\includegraphics[scale=0.25]{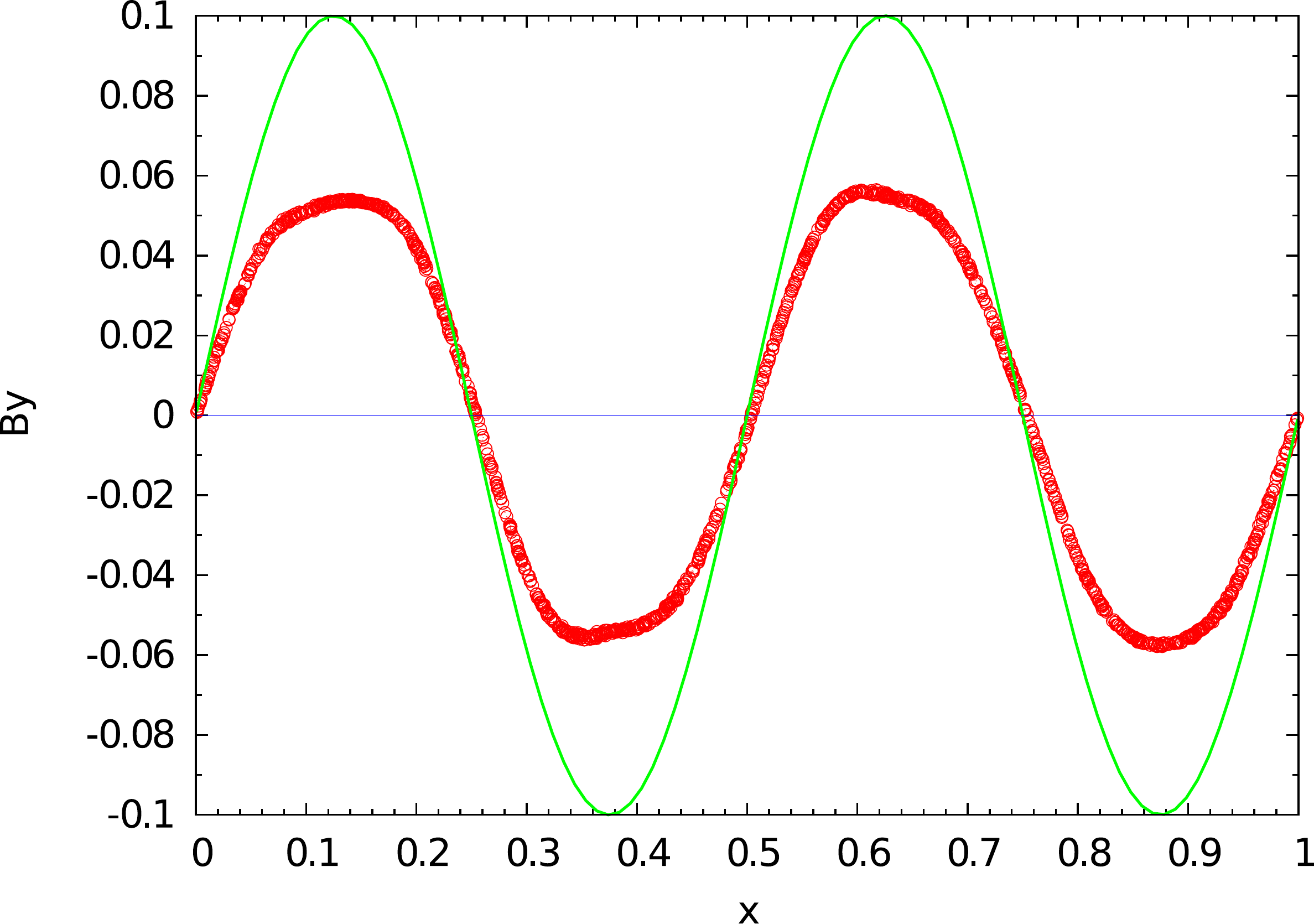}
\includegraphics[scale=0.25]{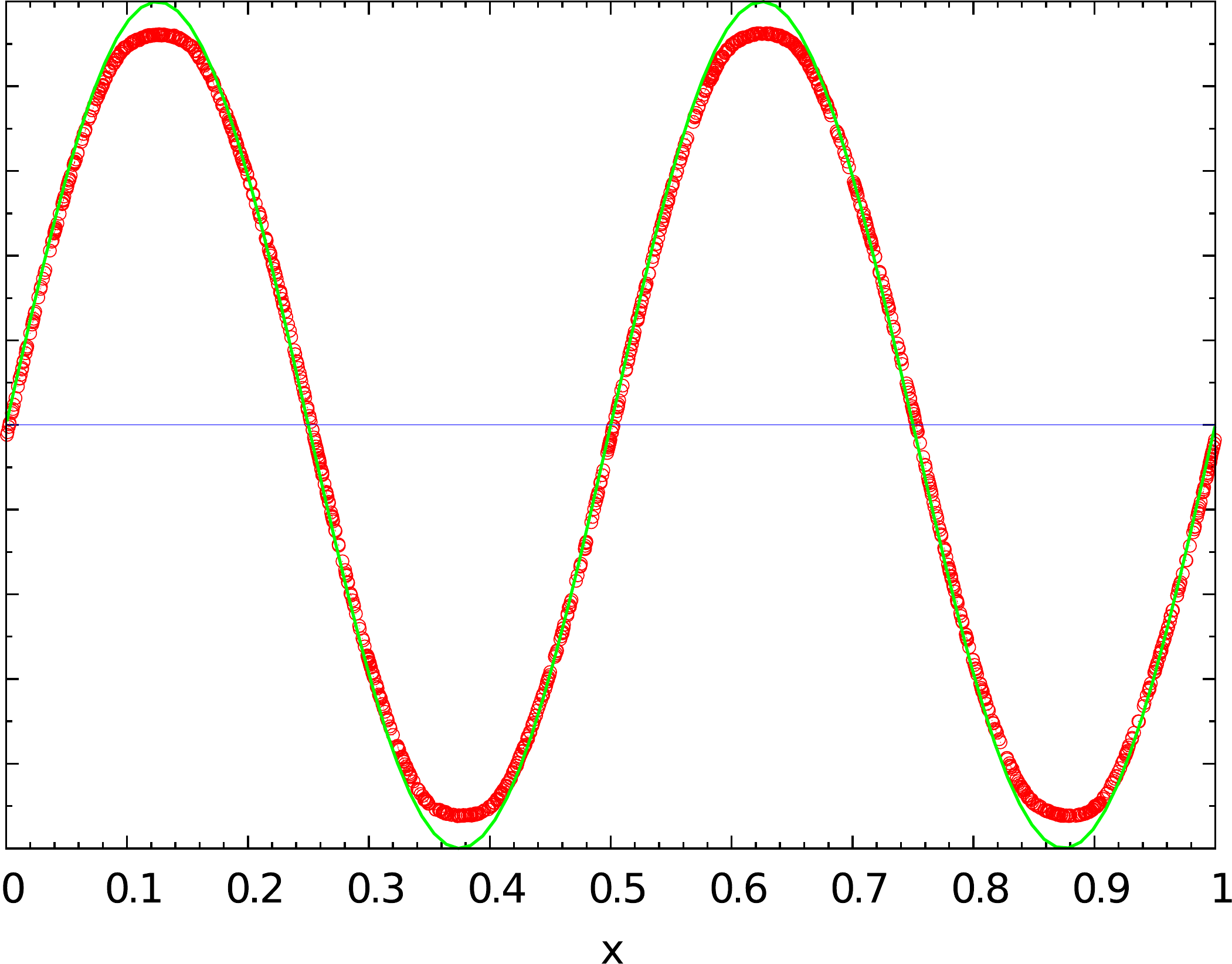}
\includegraphics[scale=0.25]{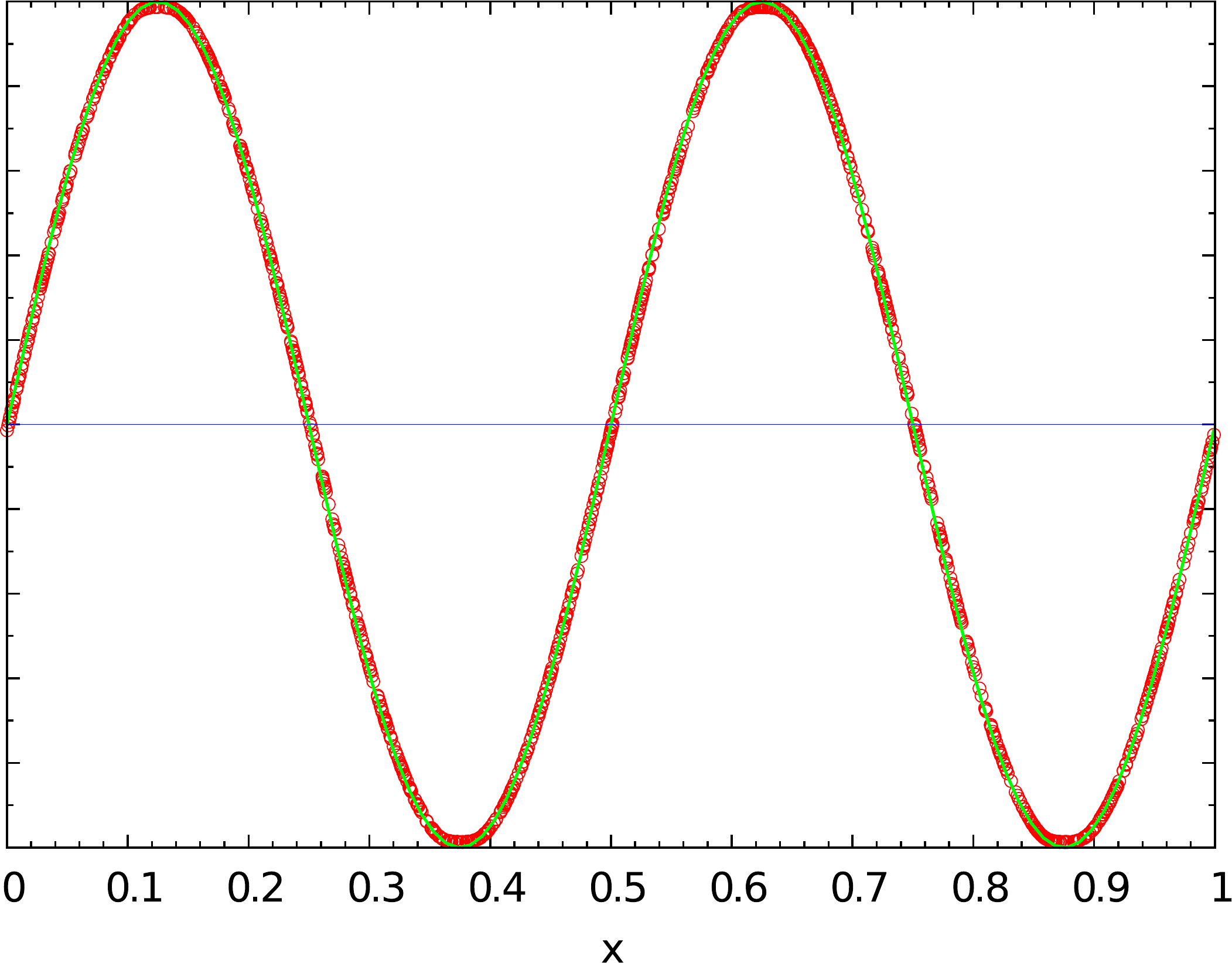} \end{center} \caption{The
circularly polarized Alfven wave after five crossings of computational domain.
The panels show the $y$-component of the magnetic field as a function of
$x$-coordinate. The solid line demonstrates exact solution, while the open
circles show the result of simulations.  In the left-most, middle and
right-most panels, the wavelength is resolveid with an average of $13$, $26$
and $52$ meshpoints respectively.} \label{fig:A1_alfven_profile} \end{figure*}
In Fig.\,\ref{fig:A1_alfven_profile} we compare the simulated results to the
analytical solution. It can be seen that lower resolution simulations have more
dissipation but do not introduce phase error in the solution. The dissipation
is not the result of the underlying Riemann-solver or a reconstruction method,
but rather that of non-linear monotonicity constraints on the linear
reconstruction model which is required for a stable description of
discontinuities. The side effect of these is the constraint is that it forces
the scheme to be first order accurate at extrema \citep{1979JCoPh..32..101V,
2011MNRAS.418.1668I}.

             \begin{figure*}[] \begin{center}
\includegraphics[scale=0.4]{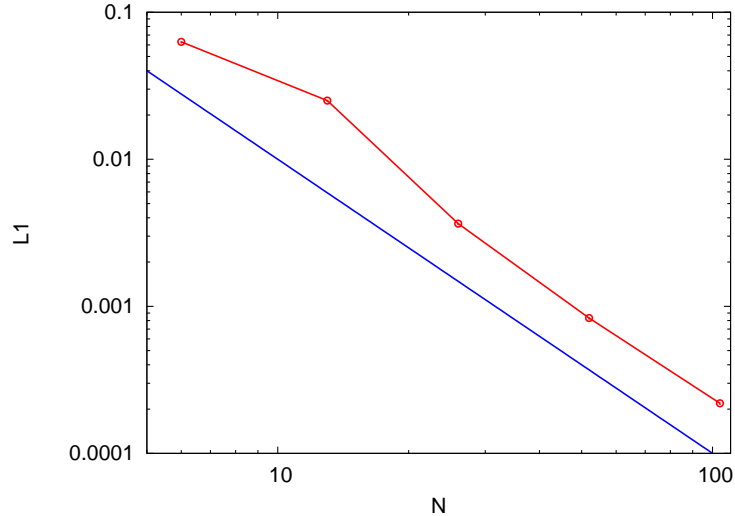} \end{center} \caption{The
$L_1$ error as a function of resolution for circularly polarized Alfven wave.
The open circles connected by the red solid line show results of simulations,
and the blue solid line is expected dependence for the second order scheme
$\propto O(N^{-2})$. The vertical axis show $L_1$ error in the solution as a
faction of number of meshpoints, $N$, per wavelength.} \label{fig:A1_alfven_L1}
\end{figure*} In Fig.\,\ref{fig:A1_alfven_L1} we show $L_1$ error of the
simulations solution as a function of the number of meshpoints per unit
wavelength. Here, the $L_1$ error is defined as $L_1=1/N \sum_i |f_i - f_{\rm
ex}|$ where sum is carried out over all $N$ mesh-cells, and $|f_i - f_{\rm
ex}|$ is absolute deviation of the value in a cells, $f_i$, from the
corresponding exact solution, $f_{\rm ex}$.  The result demonstrates that the
convergence for this problem is consistent with the second-order scheme.

                 \subsection{Magneto-rotational instability in non-stratified
cylindrical disk} In this problem we study the ability of our code to reproduce
analytical growth-rates of axisymmetric magneto-rotational instability. Our
computational domain consist of the three-dimensional non-stratified
cylindrical disk. The inner and outer radii of the disk are equal to $R=1$ and
$R=8$, and the thickness of the disk is $H=1$. We use periodic boundary
conditions in $z$ direction, and outflow boundaries at $R = 1$ and $R = 8$. The
total computational domain is a box with size $[16.6\times16.6\times 1]$.

                 We simulated three models with an average $14, 20$ and $28$
meshpoints in $z$-direction. The initial density is set to unity, and we used
isothermal equations of state with constant sound speed $c_s = 0.1$. The
gravitational potential is equal to $\phi = -1/R$, where $R = \sqrt{x^2+y^2}$,
and the initial velocity is equal to the Keplerian velocity. Initially, we set
a uniform magnetic field in $2 < R < 4$ annulus of the disk with such strength
that results in fastest growth for $n=2$ mode at $R \approx 2$. Namely we have,
$B_x = B_y = 0$ and $B_z \approx 0.055/n$, where $n=2$. In other words, at $R
\approx 2$ the fastest growing MRI mode has the wavelength $\lambda_{\rm MRI}
\approx H/2$. In this setup, the $\lambda_{\rm MRI}$ is resolved with
approximated $7$, $10$ and $14$ meshpoints in low, medium and high-resolution
simulations respectively.

                 \begin{figure*}[] \begin{center}
\includegraphics[scale=0.4]{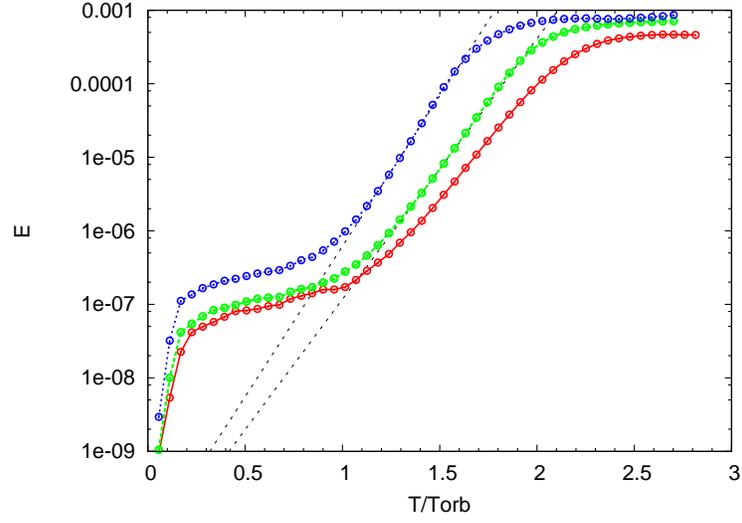} \end{center} \caption{This
figure shows radial magnetic energy in an annulus $2 < R < 2+1/42$ (vertical
axis) as a function of the number of local orbits at $R=2$ (horizontal axis).
The solid red line shows time evolution of radial magnetic energy for low
resolution simulation (on average 7 meshpoints per $\lambda_{\rm MRI}$), the
green dashed and blue dotted lines shows the results for medium (10 meshpoints
per $\lambda_{\rm MRI}$) and high resolution (14 meshpoints per $\lambda_{\rm
MRI}$). The left and right dotted lines show exponential growth with slopes
$0.75\gamma$ and $0.65\gamma$ respectively, where $\gamma = 4\pi$.}
\label{fig:A1_mri_growth} \end{figure*} In Fig.\,\ref{fig:A1_mri_growth} we
show time evolution of radial magnetic energy $E = B_r^2/2$ as a function of
the number of orbits at $R = 2$ for three different resolutions. All
simulations show exponential growth rater after approximately one orbital
period, and the low resolution simulations shows growth rate $\approx
0.6\Omega$, whereas the medium and high resolution simulations show growth rate
$\approx 0.65\Omega$ and $\approx 0.75\Omega$ respectively.

\end{document}